\documentclass[twocolumn,amsmath,amssymb,floatfix,prl,footinbib,superscriptaddress]{revtex4-1}

\usepackage{amsmath,amssymb,natbib,bm,graphicx,url,epsf,color}
\usepackage[english]{babel}
\usepackage{wasysym}
\usepackage{MnSymbol}
\usepackage{amsmath}
\usepackage{graphicx}
\usepackage{pdfpages}
\usepackage[colorinlistoftodos]{todonotes}
\usepackage[colorlinks=true, allcolors=blue]{hyperref}

\newcommand{\QH}{QH~}

\newcommand{\Vg}{V_{\mathrm{g}}}
\newcommand{\Idc}{I_{\mathrm{dc}}}

\newcommand{\Jq}{J_{\mathrm{Q}}}
\newcommand{\Jin}{J_{\mathrm{in}}}
\newcommand{\Jqe}{J_{\mathrm{Q}}^\mathrm{e}}
\newcommand{\Tc}{T_{\mathrm{c}}}

\newcommand{\DeltaTc}{\Delta T_{\mathrm{c}}}
\newcommand{\kB}{k_{\mathrm{B}}}
\newcommand{\kappazero}{\pi^2 k_{\mathrm{B}}^2/3h}

\newcommand{\sigmaxx}{\sigma_\mathrm{xx}}

\makeatletter
\AtBeginDocument{\let\LS@rot\@undefined}
\makeatother

\makeatletter
\def\NAT@spacechar{}
\makeatother

\begin{document}
\title{Heat equilibration of integer and fractional quantum Hall edge modes in graphene}

\author{G. Le Breton}
\affiliation{Universit\'e Paris-Saclay, CEA, CNRS, SPEC, 91191 Gif-sur-Yvette cedex, France
}
\author{R. Delagrange}
\affiliation{Universit\'e Paris-Saclay, CEA, CNRS, SPEC, 91191 Gif-sur-Yvette cedex, France
}
\author{Y. Hong}
\affiliation{Universit\'e Paris-Saclay, CNRS, Centre de Nanosciences et de Nanotechnologies (C2N), 91120 Palaiseau, France
}
\author{M. Garg}
\affiliation{Universit\'e Paris-Saclay, CEA, CNRS, SPEC, 91191 Gif-sur-Yvette cedex, France
}
\author{K. Watanabe}
\affiliation{National Institute for Materials Science, Tsukuba, Japan
}
\author{T. Taniguchi}
\affiliation{National Institute for Materials Science, Tsukuba, Japan
}
\author{R. Ribeiro-Palau}
\affiliation{Universit\'e Paris-Saclay, CNRS, Centre de Nanosciences et de Nanotechnologies (C2N), 91120 Palaiseau, France
}
\author{P. Roulleau}
\affiliation{Universit\'e Paris-Saclay, CEA, CNRS, SPEC, 91191 Gif-sur-Yvette cedex, France
}
\author{P. Roche}
\affiliation{Universit\'e Paris-Saclay, CEA, CNRS, SPEC, 91191 Gif-sur-Yvette cedex, France
}
\author{F.D. Parmentier}
\affiliation{Universit\'e Paris-Saclay, CEA, CNRS, SPEC, 91191 Gif-sur-Yvette cedex, France
}

\date{\today}

\begin{abstract}
Hole-conjugate states of the fractional quantum Hall effect host counter-propagating edge channels which are thought to exchange charge and energy. These exchanges have been the subject of extensive theoretical and experimental works; in particular, it is yet unclear if the presence of integer quantum Hall edge channels stemming from fully filled Landau levels affects heat equilibration along the edge. In this letter, we present heat transport measurements in quantum Hall states of graphene demonstrating that the integer channels can strongly equilibrate with the fractional ones, leading to markedly different regimes of quantized heat transport that depend on edge electrostatics. Our results allow for a better comprehension of the complex edge physics in the fractional quantum Hall regime.
\end{abstract}

\maketitle

\begin{figure*}[ht]
\centering

\includegraphics[width=0.94\textwidth]{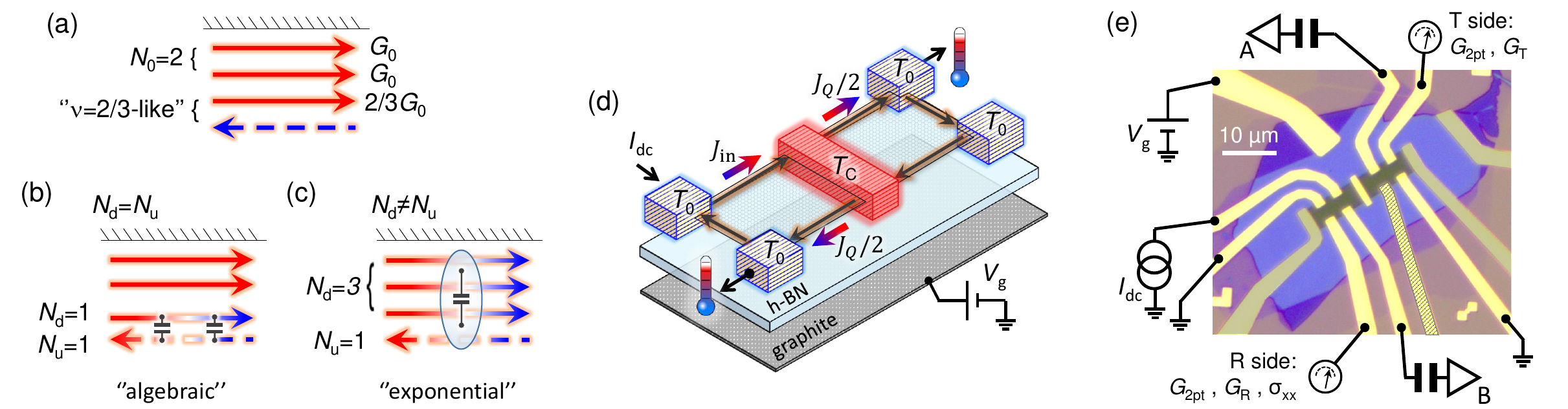}
\caption{\label{fig1-sample} $\nu=8/3$ edge structure, without heat equilibration \textbf{(a),} with heat equilibration between fractional modes only \textbf{(b),} and between all modes \textbf{(c)}. Full/dashed arrows: charged/neutral modes, with their chirality. Arrow colour: temperature gradient in the presence of a temperature bias: hot source (red) on the left, cold source (blue) on the right. \textbf{(d),} Schematic representation of the experiment, with cold electrodes (blue) at $T_0$ and the metallic island (red) at $\Tc$. Red arrows: chiral edge channels, gradient-coloured arrows: heat flows going in ($\Jin$) and out of ($\Jq$) the island. \textbf{(e),} Optical micrograph of the sample, with the experimental wiring. The encapsulated graphene flake is shown in green. The greyed out electrodes are left floating, and the hatched electrode is used as a current feed in cooldown 2.}
\end{figure*}

The fractional quantum Hall (QH) effect emerges when a two-dimensional electron system is subjected to a strong perpendicular magnetic field $B$ such that the filling factor $\nu=n_\mathrm{e}h/eB$ takes fractional values ($n_\mathrm{e}$ is the carrier density, $h$ Planck's constant, and $-e$ the electron charge). For hole-conjugate states, it takes the form $\nu=N_0+1-p/q$, with $N_0$ the integer part of $\nu$ corresponding to fully filled Landau levels (LLs), $q$ an odd number, and $p$ such that $p/q<1/2$. The edge structure for such states has been the subject of more than 30 years of research, originally focused on $\nu=2/3$ (such that $N_0=0$ and $p/q=1/3$)~\cite{Beenakker1990,MacDonald1990}. Some of the earlier works proposed the $\nu=2/3$ edge to be composed of one downstream channel with integer electrical conductance $G_0=e^2/h$ along with one \emph{upstream} channel with fractional conductance $-1/3\times G_0$~\cite{MacDonald1990}. It was later proposed that inter-channel interactions and disorder-assisted charge tunneling between the downstream and upstream channels radically change that structure. Strong interactions give rise to a downstream charged mode with fractional electrical conductance $2/3\times G_0$ and one upstream \emph{neutral} mode which only carries heat in the direction opposite to that of charge transport~\cite{Kane1994}. This \emph{charge equilibration} was then generalized to other fractions~\cite{Kane1995,Kane1997}. Importantly, depending on $\nu$ the numbers of downstream fractional-charged modes and upstream neutral modes are not necessarily equal. Neutral modes were first observed in 2010~\cite{Bid2010}, then extensively investigated using shot noise~\cite{Dolev2011,Gross2012,Inoue2014} and local thermometry~\cite{Venkatachalam2012} measurements.

Recently, the question of \emph{heat equilibration} between neutral and charged modes has been the center of a growing number of works, both experimental~\cite{Banerjee2017,Banerjee2018,Srivastav2021,Melcer2022} and theoretical~\cite{Protopopov2017,Nosiglia2018,Ma2019,Spanslatt2019,Aharon-Steinberg2019,Ma2020}. While most experiments confirm a charge equilibration (see \textit{e.g.} Refs.~\cite{Cohen2019,Lafont2019} for notable exceptions), heat equilibration is much less universal. In gallium arsenide (GaAs) based 2-dimensional electron gases, partial to full heat equilibration was first reported at $\nu=2/3,3/5,4/7$~\cite{Banerjee2017}, and $8/3$~\cite{Banerjee2018}; however, a recent experiment showed an absence of heat equilibration at $\nu=2/3$ even for large ($>300~\mu$m) lengths~\cite{Melcer2022}. Experiments in graphene reported no heat equilibration at $\nu=5/3$ and $8/3$~\cite{Srivastav2021} over a few microns scale, and, very recently, the observation of a temperature-induced heat equilibration at $\nu=2/3$ and $3/5$~\cite{Srivastav2022}. This diversity of observations is currently understood by the facts that the charge and heat equilibration lengths can be largely different depending on the coupling between the counter-propagating edge modes~\cite{Srivastav2021}, and that the ratio between the number of coupled downstream modes $N_\mathrm{d}$ and upstream modes $N_\mathrm{u}$ strongly affects the equilibration. Namely, for states with $N_\mathrm{d}=N_\mathrm{u}$ (\textit{e.g.} $\nu=2/3$), heat equilibration is predicted to have slow algebraic length dependence~\cite{Aharon-Steinberg2019,Spanslatt2019,Srivastav2022}, and is not observed at low temperature, even at large length scales~\cite{Melcer2022}. On the contrary, for $N_\mathrm{d}\neq N_\mathrm{u}$ it should be exponentially fast~\cite{Aharon-Steinberg2019,Spanslatt2019,Srivastav2022}. However, it is still unclear whether, for fractional $\nu>1$, the $N_0$ integer edge channels (ECs) stemming from the fully filled LLs participate in the heat equilibration along the edge~\cite{Ma2019}. If so, one should include them in the $N_\mathrm{d}$ downstream modes, which can lead to $N_\mathrm{d}\neq N_\mathrm{u}$ in states where $\nu=N_0+2/3$, radically changing heat equilibration.

We addressed this question by probing heat transport in graphene at filling factor $\nu=8/3$. Fig.~\ref{fig1-sample}a shows its edge structure, with $N_0=2$ integer ECs stemming from the fully filled zeroth LL, and a $\nu=2/3$-like pair of counterpropagating fractional edge modes~\cite{Ma2019}. The upstream mode can either exchange heat with only the fractional downstream mode (Fig.~\ref{fig1-sample}b). This 'algebraic' case is similar to $\nu=2/3$, with $N_\mathrm{d}=N_\mathrm{u}=1$, such that no heat equilibration is expected at low temperature and short/moderate lengths~\cite{Srivastav2021,Melcer2022,Srivastav2022}. Conversely, the upstream mode can exchange heat with all downstream channels (Fig.~\ref{fig1-sample}c), such that $N_\mathrm{d}=N_0+1=3$ and $N_\mathrm{u}=1$, implying a much more efficient heat equilibration. This difference is directly reflected in the heat flow, affecting the number $N$ of effective ballistic heat transport channels~\cite{Kane1997,Aharon-Steinberg2019,Spanslatt2019,Srivastav2021}. In the non-equilibrated (algebraic) case, all downstream and upstream modes are ballistic and contribute independently, yielding $N=4$. In the fully equilibrated case, the upstream mode suppresses heat transport down to $N=2$.

Fig.~\ref{fig1-sample}d shows our experimental principle. It was first demonstrated in GaAs in the integer \QH regime~\cite{Jezouin2013a}, and later applied to the fractional \QH effect~\cite{Banerjee2017,Banerjee2018}. Recent experiments~\cite{Srivastav2019,Srivastav2021,Srivastav2022} have extended it to graphene. A 2-dimensional electron gas (here in graphene) is divided in two regions electrically connected by a floating metallic island, highlighted in red in Fig.~\ref{fig1-sample}b. A perpendicular magnetic field $B$ allows reaching the \QH regime, with equal $\nu$ in both regions. The dc electrical current $\Idc$ is applied to one of the cold electrodes (in blue in Fig.~\ref{fig1-sample}d), and flows downstream via the ECs (red lines in Fig.~\ref{fig1-sample}d) to the island. The latter evenly splits the current between the outgoing ECs in the two regions, resulting in a net Joule power directly dissipated into the island $\Jin=\Idc^2/(4\nu G_0)$~\cite{Jezouin2013a,Srivastav2019,SM}. This induces an increase in the electron temperature $\Tc$ of the island, while all other electrodes remain at base electron temperature $T_0$. The input heat flow $\Jin$ is evacuated from the island through the outgoing ECs on both sides of the island, each side carrying half of the outgoing heat flow, $\Jq/2$. Each ballistic channel carries a quantum-limited heat flow $\Jqe=0.5\kappa_0(\Tc^2-T_0^2)$~\cite{Pendry1983,Rego1999,Jezouin2013a}, with $\kappa_0=\kappazero$ ($\kB$ is Boltzmann's constant). Neglecting other contributions (\textit{e.g.} coupling to phonons, see below), the heat balance simply reads $\Jq=\Jin$, hence

\begin{equation}
\Jq = \frac{1}{4\nu G_0}\Idc^2 = 2 N \frac{\kappa_0}{2}(\Tc^2-T_0^2),
\label{eq:heatbalance}
\end{equation}

where $N$ is the number of ballistic heat-carrying channels flowing out of each side of the island (the total number thus being $2N$). For integer \QH states, $N$ equals the filling factor $\nu$. For hole-conjugate fractional \QH states, $N$ reflects the heat equilibration along the edge, as detailed above. $N$ can be directly extracted by measuring the temperature $\Tc$ and comparing it to the input heat flow according to Eq.~\ref{eq:heatbalance}.

\begin{figure*}[ht!]
\centering
\includegraphics[width=0.9\textwidth]{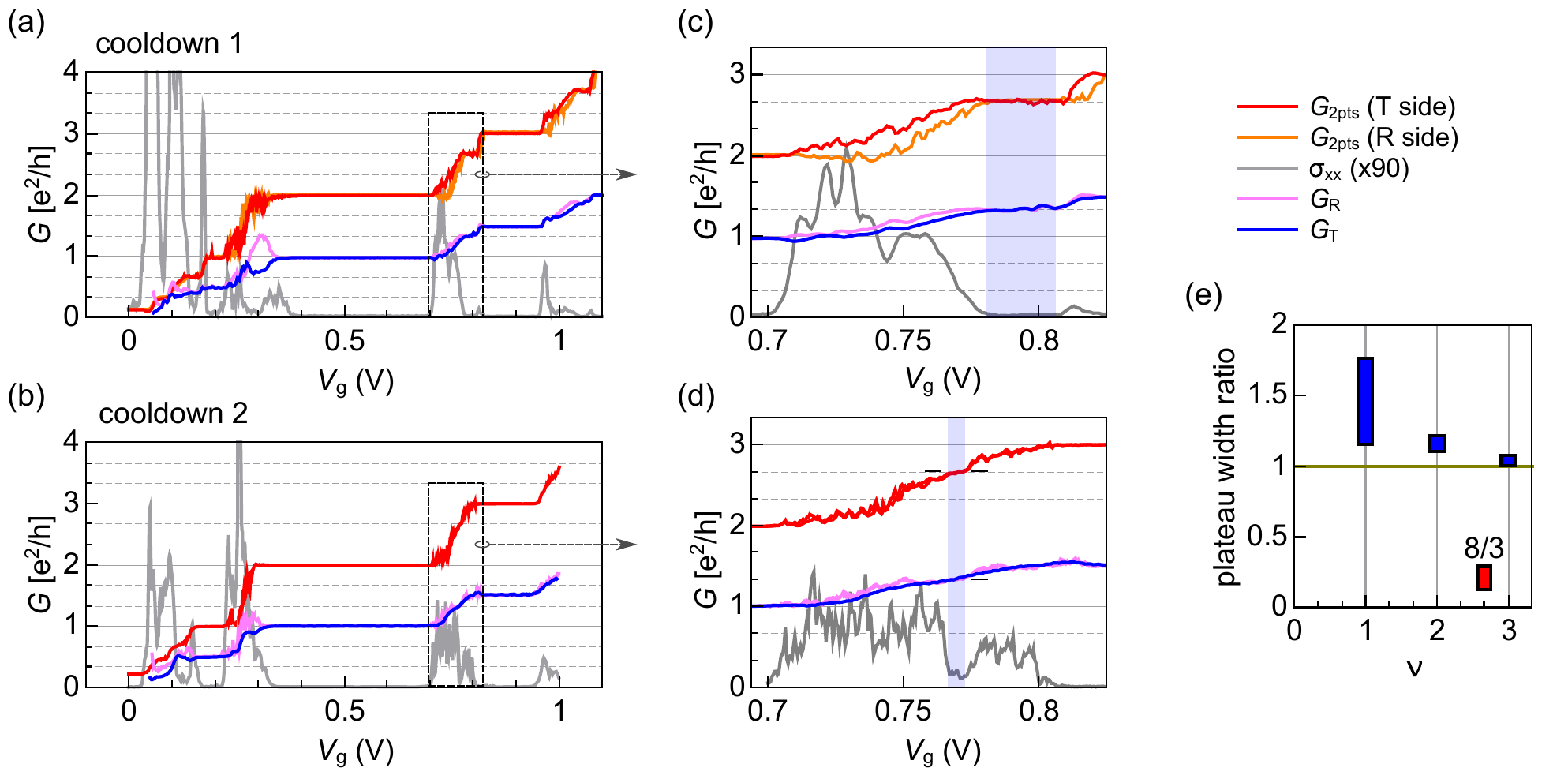}
\caption{\label{fig2-GvsVg} Conductances versus $\Vg$, measured at $B=7~$T and $T=10~$mK for \textbf{(a)} cooldown 1 and \textbf{(b)} cooldown 2. Red (resp. orange): 2-point conductance $G_{2\mathrm{pt}}$ of the transmitted (resp. reflected) side (see legend on the upper right corner). Grey: longitudinal-like conductivity $\sigmaxx$. Lavender: reflected transconductance $G_\mathrm{R}$. Blue: transmitted transconductance $G_\mathrm{T}$. 
\textbf{c} and \textbf{d}: Zooms on the $\nu=2\rightarrow 3$ transition (dashed rectangles in \textbf{(a)} and \textbf{(b)}). The $\nu=8/3$ region is highlighted in blue. Horizontal ticks in \textbf{(d)}: guides for the eyes at the expected values of $G_{2\mathrm{pt}}$ and $G_\mathrm{T/R}$. \textbf{(e),} Ratio between the $\Vg$ widths of the $\nu=1,2,3$ (blue) and $\nu=8/3$ (red) plateaus between cooldown 2 and cooldown 1. The maximum of the bars corresponds to the ratio of the widths extracted from $G_{2\mathrm{pt}}$, and the minimum to the one extracted from $\sigmaxx$.}
\end{figure*}

Fig.~\ref{fig1-sample}e shows our implementation in a hexagonal boron nitride (h-BN)-encapsulated monolayer graphene sample. The charge carrier type and density are tuned using a graphite back gate upon which the voltage $\Vg$ is applied. The Ti/Au metallic island has dimensions $6.8~\mu$m~$\times1.25~\mu$m~$\times100~$nm, its distance to the closest electrodes is $\sim2.5~\mu$m, and the width of the device is $\sim5~\mu$m. ECs flowing out of the two sides of the island, denoted 'reflected' (R) and 'transmitted' (T) with respect to the current feed, connect to measurement electrodes in this order: noise, low-frequency conductance, cold ground. We characterize charge transport by measuring the 2-point differential conductances $G_{2\mathrm{pt}}=(dV_\mathrm{R,T}/d\Tilde{I}_\mathrm{R,T})^{-1}$ ($\Tilde{I}_\mathrm{R,T}$ are the currents directly applied to the measurement contacts on the R and T side), the differential transmitted and reflected transconductances $G_\mathrm{R,T}= (dV_\mathrm{R,T}/d \Idc)^{-1}$ probing current redistribution at the island, and the 'longitudinal-like' differential conductance $\sigmaxx=G_0^2\times dV_\mathrm{R}/d\Tilde{I}_\mathrm{T}$. The latter vanishes for well-defined \QH states because the chiral paths connecting the conductance measurement electrodes are interrupted by a cold ground. The island's electron temperature increase $\DeltaTc$ induces current fluctuations $\Delta S= \nu G_0 \kB \DeltaTc$ in the ECs flowing out of the island, that we detect through two independent noise measurement lines on each side (A and B in Fig.~\ref{fig1-sample}e).

We present measurements in two consecutive cooldowns of the same device. All connections were kept identical, except for the current feed which was swapped between the R side in cooldown 1 (CD1) and the T side in cooldown 2 (CD2).

\begin{figure}[ht]
\centering
\includegraphics[width=0.4\textwidth]{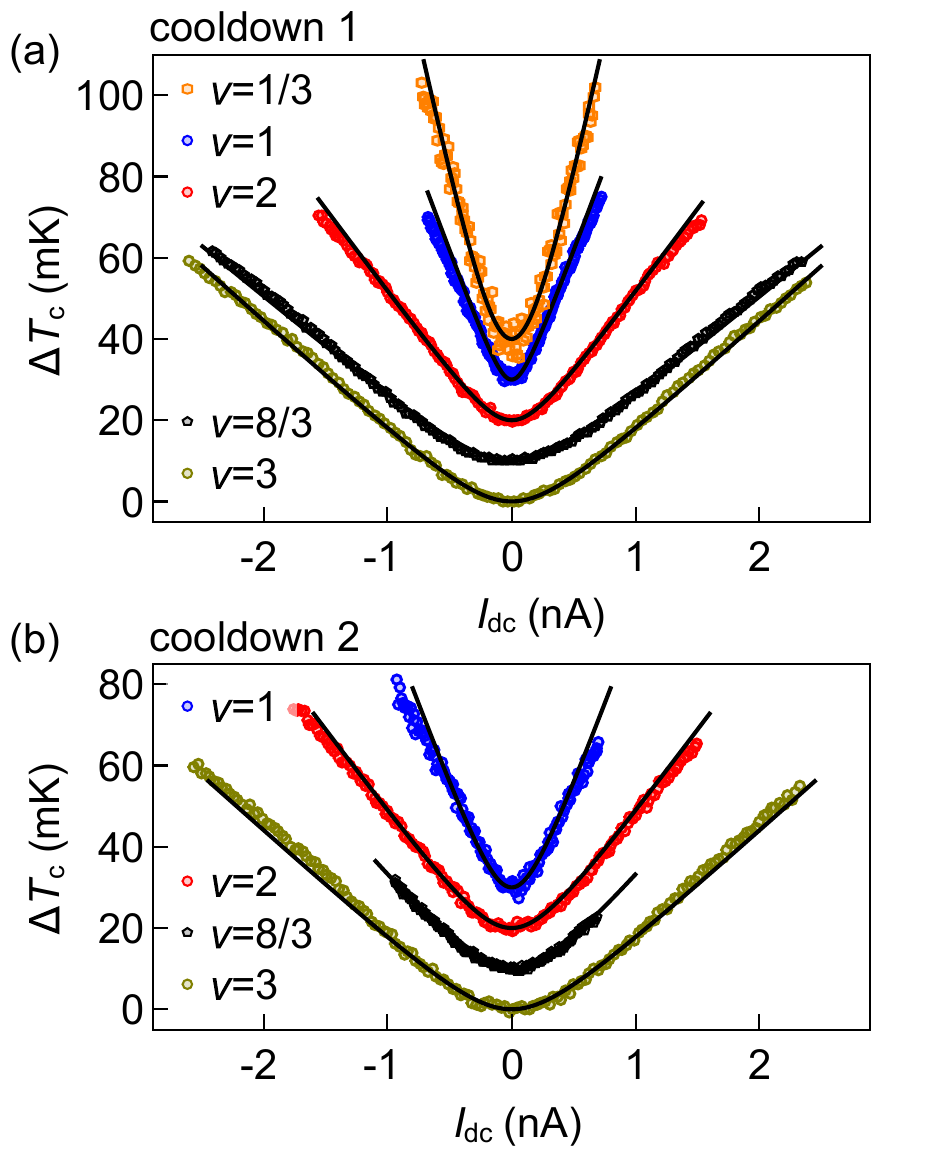}
\caption{\label{fig3-TvsI} $\Delta T_\mathrm{c}$ versus $I_\mathrm{dc}$, for cooldown 1 \textbf{(a)} and 2 \textbf{(b)}. Symbols: experimental data (orange $\varhexagon$~: $\nu=1/3$, blue $\circ$~: $\nu=1$, red $\circ$~: $\nu=2$, black $\pentagon$~: $\nu=8/3$, dark yellow $\circ$~: $\nu=3$). Black lines: fits (see text). Data and fits corresponding to each $\nu$ are vertically shifted by increments of $10~$mK for clarity.
}
\end{figure}

Fig.~\ref{fig2-GvsVg} shows conductance measurements as a function of the gate voltage, obtained at $B=7~$T, for CD1 (Fig.~\ref{fig2-GvsVg}a,c) and CD2 (Fig.~\ref{fig2-GvsVg}b,d). Well-defined \QH states, at both integer and fractional $\nu$, are characterized by quantized plateaus in the 2-point conductances $G_{2\mathrm{pt}}=\nu G_0$, along with a vanishing $\sigmaxx$. On most of these plateaus (except notably on $\nu=1/3$ in CD2), $G_\mathrm{R}$ and $G_\mathrm{T}$ have equal values, quantized to $0.5\times \nu G_0$, indicating chiral charge transport and near-ideal current redistribution at the island. Figs.~\ref{fig2-GvsVg}c,d show a zoom on the $\nu=2\rightarrow 3$ transition. The width of the $\nu=8/3$ plateau is strongly reduced in CD2, with a non-zero local minimum in $\sigmaxx$. Nevertheless, chirality and current redistribution are still preserved: both transconductances are equal, with a plateau at half-quantized value $0.5\times 8/3 G_0$~\cite{SM}. We superimpose two traces in Fig.~\ref{fig2-GvsVg}d, illustrating the reproducibility of this feature. 

Fig.~\ref{fig2-GvsVg}e plots the ratio between the $\Vg$ widths of the plateaus between CD2 and CD1. These spans can be extracted from the quantized $G_{2\mathrm{pt}}$, or from the minima in $\sigmaxx$; either show that for CD2, all integer \QH plateaus are wider while $\nu=8/3$ is markedly narrower. The $\Vg$ position of each plateau can be similarly extracted~\cite{SM}; we observe a systematic shift towards more negative $\Vg$ at CD2, corresponding to an increased intrinsic electron doping $\Delta n_e\approx1.7\times 10^{10}~$cm$^{-2}$.

Thermal measurements were performed for each \QH state in which the chirality and current redistribution criteria are enforced: $\nu=\{1/3,1,2,8/3,3\}$ for CD1, and $\nu=\{1,2,8/3,3\}$ for CD2. We use auto- and cross-correlations of the two noise lines to extract $\DeltaTc$ from spurious noise contributions~\cite{Sivre2019,SM}. Fig.~\ref{fig3-TvsI} shows $\DeltaTc$ measured as a function of the dc current $\Idc$. All filling factors display the same qualitative behavior, in agreement with Eq.~\ref{eq:heatbalance}, where $\DeltaTc$ increases linearly beyond a thermal rounding at low $\Idc$. Following Eq.~\ref{eq:heatbalance}, the slope only depends on $1/\sqrt{N\times\nu}$. The data show similar slopes for both cooldowns, except $\nu=8/3$, with a markedly higher slope at CD2. This observation, discussed in detail below, is the main result of our work. Note that for this last dataset, we kept $\Idc$ small, as both auto-correlations became different above $ |\Idc |\approx1~$nA~\cite{SM}.

We compared our data to Eq.~\ref{eq:heatbalance}, assuming negligible electron-phonon cooling in the island. This is reasonable given its small volume and the very low temperatures~\cite{SM}; previous experiments in graphene~\cite{Srivastav2019, Srivastav2021, Srivastav2022}, withsimilar dimensions, also reported negligible electron-phonon cooling. Eq.~\ref{eq:heatbalance} appears as black lines in Fig.~\ref{fig3-TvsI}. The number of ballistic heat carrying modes $N$ is fixed to its expected value ($N=\nu$ for integer \QH states, see below for $\nu=8/3$), yielding an excellent agreement with the slope of the data. The thermal rounding is reproduced by adjusting the base electron temperature $T_0$ for each $\nu$. These extracted $T_0$ match with the equilibrium Johnson-Nyquist noise measured at $\Idc=0$~\cite{SM}. For CD1, $T_0\approx12~$mK (for a fridge base temperature of $8.7~$mK), with the notable exception of $\nu=1/3$, where $T_0\approx42~$mK. For CD2, $T_0\approx 15~$mK, except at $\nu=1$, where $T_0\approx20~$mK. We attribute those variations, particularly the increase at lower $\nu$, to mechanical vibrations~\cite{Iftikhar2016}. Additional analysis (\textit{e.g.} heat Coulomb blockade effects~\cite{Sivre2018}), described in~\cite{SM}, yields a reasonable uncertainty on the extracted $T_0$ of about $\pm3~$mK, translating into a typical uncertainty on $N$ of about $\pm0.1$.

\begin{figure}[ht]
\centering
\includegraphics[width=0.4\textwidth]{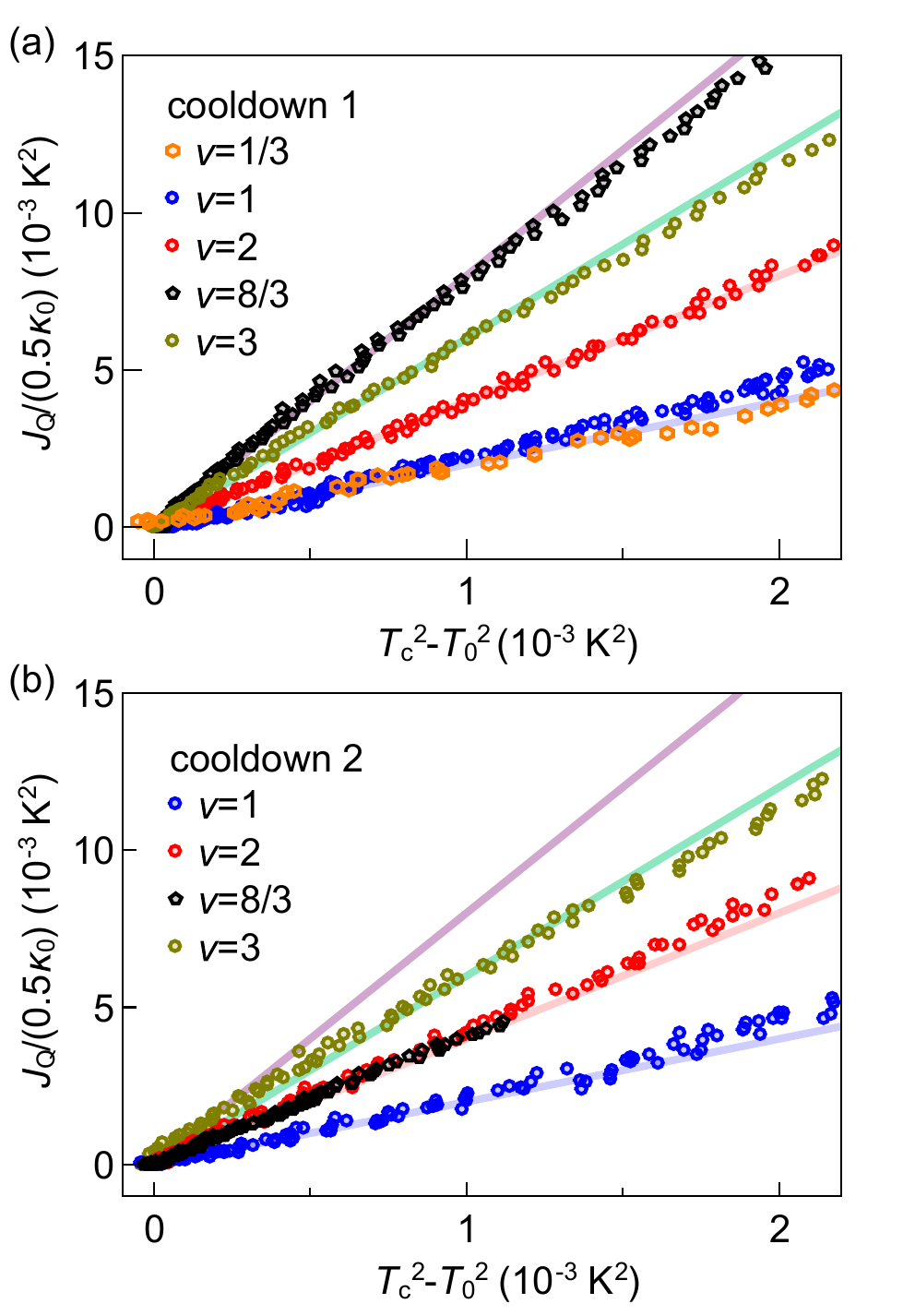}
\caption{\label{fig4-JqvsT2} Heat flow $\Jq$ in units of $\kappa_0/2$ versus $\Tc^2-T_0^2$, for cooldown 1 \textbf{(a)} and 2 \textbf{(b)}. Symbols: experimental data corresponding to the one in Fig.~\ref{fig3-TvsI} (orange $\varhexagon$~: $\nu=1/3$, blue $\circ$~: $\nu=1$, red $\circ$~: $\nu=2$, black $\pentagon$~: $\nu=8/3$, dark yellow $\circ$~: $\nu=3$). Lines: theoretical predictions for the quantized heat flow carried by $2N$ ballistic channels, with $N=1$ (light blue), $N=2$ (pink), $N=3$ (light green) and $N=4$ (lavender).}
\end{figure}

The relation between the slopes in $\DeltaTc(\Idc)$ and quantized heat transport appears clearly when replotting the data in terms of the total heat flow leaving the island $\Jq=\Idc^2/(4\nu G_0)$ as a function of $(\Tc^2-T_0^2)$. This is shown in Fig.~\ref{fig4-JqvsT2}, with $T_0$ extracted from the above procedure, such that the data ($\Jq$ is plotted in units of $0.5\kappa_0$) naturally fall onto lines with integer slope $2N$. The representation of Fig.~\ref{fig4-JqvsT2} shows heat transport properties of each filling factor regardless of charge transport. As a striking example, the data at $\nu=1/3$ and $\nu=1$ fall onto the same $N=1$ line, demonstrating that a fractional and an integer ECs carry the same universally quantized heat flow $\pi^2 k_{\mathrm{B}}^2/6h(\Tc^2-T_0^2)$~\cite{Pendry1983,Rego1999}, previously reported in GaAs~\cite{Banerjee2017} and graphene~\cite{Srivastav2019,Srivastav2022}. Fig.~\ref{fig4-JqvsT2} emphasizes the remarkable difference between both cooldowns for $\nu=8/3$. On the one hand, in CD1 we observe a quantized heat flow with $N=4$ channels, corresponding to non-equilibrated ballistic heat transport through all downstream charged modes and the upstream neutral mode. This is consistent with recent results in bilayer graphene~\cite{Srivastav2021}. On the other hand, for CD2, $\nu=8/3$ falls on top of $\nu=2$, corresponding to a quantized heat flow with $N=2$ channels. This unambiguously signals a strong heat equilibration at $\nu=8/3$, corresponding to the upstream neutral mode fully equilibrating with both integer and fractional downstream charged modes.

The increase of disorder and doping revealed in Fig.~\ref{fig2-GvsVg} thus leads to a large change in the thermal conductance of $\nu=8/3$ between the two cooldowns. This strongly suggests that heat equilibration in CD2 is exponential, confirming the fact that the integer ECs have to be considered in this equilibration process. Microscopically, the increase electron doping, likely stemming from charged impurities adsorbed at the surface of the sample while it was exposed to ambient air during thermal cycling, can favour efficient equilibration. Indeed, these impurities locally increase the electron density in the vicinity of the edge, resulting in a sharper edge confinement potential which increases the coupling between the more closely packed ECs. Not only can the increased coupling drive the heat equilibration in an exponential regime, but it can also drastically affect the characteristic length~\cite{Srivastav2022}, further favoring equilibration. Even though spatially separated~\cite{Ma2019}, the fractional and integer channels can thus be strongly coupled; interestingly, this can be related to recent observations of charge tunneling between integer channels at $\nu=3$ in graphene~\cite{Amet2014,Wei2017}.

Finally, the small, non-zero $\sigmaxx$ measured in CD2 raises the question whether our observations stem from bulk heat transport. It is unlikely, as this would effectively increase $N$ rather than diminish it~\cite{SM}.

In conclusion, we have observed the two opposite regimes of heat equilibration on the edge in the fractional \QH regime, suggesting that exponential heat equilibration can occur at $\nu=8/3$. Our result demonstrate the crucial importance of considering all downstream modes in the heat equilibration, particularly the integer ECs copropagating along the fractional edge modes. This is likely to impact experiments realizing new quantum circuits based on the non-trivial statistics of fractional \QH states at $\nu>1$.

\begin{acknowledgments}

We warmly thank C. Altimiras, A. Zhang, F. Pierre, A. Anthore, D. Kovrizhin, M. O. Goerbig, J. Splettstoesser and C. Sp\r{a}nsl\"{a}tt for enlightening discussions, as well as P. Jacques for precious technical assistance. This work was funded by the ERC (ERC-2018-STG \textit{QUAHQ}), and by the ''Investissements d'Avenir'' LabEx PALM (ANR-10-LABX-0039-PALM). R.R.-P. and Y.H. acknowledge financial support form the ERC (ERC-2019-STG \textit{TWISTRONICS}). P. Roulleau acknowledges financial support form the ERC (ERC-2015-STG \textit{COHEGRAPH}).
\end{acknowledgments}

\nocite{Kim2009}
\nocite{Ribeiro-Palau2019}
\nocite{Flor2022}


\begin{thebibliography}{33}%
\makeatletter
\providecommand \@ifxundefined [1]{%
 \@ifx{#1\undefined}
}%
\providecommand \@ifnum [1]{%
 \ifnum #1\expandafter \@firstoftwo
 \else \expandafter \@secondoftwo
 \fi
}%
\providecommand \@ifx [1]{%
 \ifx #1\expandafter \@firstoftwo
 \else \expandafter \@secondoftwo
 \fi
}%
\providecommand \natexlab [1]{#1}%
\providecommand \enquote  [1]{``#1''}%
\providecommand \bibnamefont  [1]{#1}%
\providecommand \bibfnamefont [1]{#1}%
\providecommand \citenamefont [1]{#1}%
\providecommand \href@noop [0]{\@secondoftwo}%
\providecommand \href [0]{\begingroup \@sanitize@url \@href}%
\providecommand \@href[1]{\@@startlink{#1}\@@href}%
\providecommand \@@href[1]{\endgroup#1\@@endlink}%
\providecommand \@sanitize@url [0]{\catcode `\\12\catcode `\$12\catcode
  `\&12\catcode `\#12\catcode `\^12\catcode `\_12\catcode `\%12\relax}%
\providecommand \@@startlink[1]{}%
\providecommand \@@endlink[0]{}%
\providecommand \url  [0]{\begingroup\@sanitize@url \@url }%
\providecommand \@url [1]{\endgroup\@href {#1}{\urlprefix }}%
\providecommand \urlprefix  [0]{URL }%
\providecommand \Eprint [0]{\href }%
\providecommand \doibase [0]{http://dx.doi.org/}%
\providecommand \selectlanguage [0]{\@gobble}%
\providecommand \bibinfo  [0]{\@secondoftwo}%
\providecommand \bibfield  [0]{\@secondoftwo}%
\providecommand \translation [1]{[#1]}%
\providecommand \BibitemOpen [0]{}%
\providecommand \bibitemStop [0]{}%
\providecommand \bibitemNoStop [0]{.\EOS\space}%
\providecommand \EOS [0]{\spacefactor3000\relax}%
\providecommand \BibitemShut  [1]{\csname bibitem#1\endcsname}%
\let\auto@bib@innerbib\@empty
%</preamble>
\bibitem [{\citenamefont {Beenakker}(1990)}]{Beenakker1990}%
  \BibitemOpen
  \bibfield  {author} {\bibinfo {author} {\bibfnamefont {C.~W.~J.}\
  \bibnamefont {Beenakker}},\ }\href {\doibase 10.1103/PhysRevLett.64.216}
  {\bibfield  {journal} {\bibinfo  {journal} {Physical Review Letters}\
  }\textbf {\bibinfo {volume} {64}},\ \bibinfo {pages} {216} (\bibinfo {year}
  {1990})}\BibitemShut {NoStop}%
\bibitem [{\citenamefont {MacDonald}(1990)}]{MacDonald1990}%
  \BibitemOpen
  \bibfield  {author} {\bibinfo {author} {\bibfnamefont {A.~H.}\ \bibnamefont
  {MacDonald}},\ }\href {\doibase 10.1103/PhysRevLett.64.220} {\bibfield
  {journal} {\bibinfo  {journal} {Physical Review Letters}\ }\textbf {\bibinfo
  {volume} {64}},\ \bibinfo {pages} {220} (\bibinfo {year} {1990})}\BibitemShut
  {NoStop}%
\bibitem [{\citenamefont {Kane}\ \emph {et~al.}(1994)\citenamefont {Kane},
  \citenamefont {Fisher},\ and\ \citenamefont {Polchinski}}]{Kane1994}%
  \BibitemOpen
  \bibfield  {author} {\bibinfo {author} {\bibfnamefont {C.~L.}\ \bibnamefont
  {Kane}}, \bibinfo {author} {\bibfnamefont {M.~P.~A.}\ \bibnamefont {Fisher}},
  \ and\ \bibinfo {author} {\bibfnamefont {J.}~\bibnamefont {Polchinski}},\
  }\href {\doibase 10.1103/PhysRevLett.72.4129} {\bibfield  {journal} {\bibinfo
   {journal} {Physical Review Letters}\ }\textbf {\bibinfo {volume} {72}},\
  \bibinfo {pages} {4129} (\bibinfo {year} {1994})}\BibitemShut {NoStop}%
\bibitem [{\citenamefont {Kane}\ and\ \citenamefont {Fisher}(1995)}]{Kane1995}%
  \BibitemOpen
  \bibfield  {author} {\bibinfo {author} {\bibfnamefont {C.~L.}\ \bibnamefont
  {Kane}}\ and\ \bibinfo {author} {\bibfnamefont {M.~P.~A.}\ \bibnamefont
  {Fisher}},\ }\href
  {https://journals.aps.org/prb/pdf/10.1103/PhysRevB.51.13449} {\bibfield
  {journal} {\bibinfo  {journal} {Physical Review B}\ }\textbf {\bibinfo
  {volume} {51}},\ \bibinfo {pages} {13449} (\bibinfo {year} {1995})}\BibitemShut
  {NoStop}%
\bibitem [{\citenamefont {Kane}\ and\ \citenamefont {Fisher}(1997)}]{Kane1997}%
  \BibitemOpen
  \bibfield  {author} {\bibinfo {author} {\bibfnamefont {C.~L.}\ \bibnamefont
  {Kane}}\ and\ \bibinfo {author} {\bibfnamefont {M.~P.~A.}\ \bibnamefont
  {Fisher}},\ }\href {\doibase 10.1103/PhysRevB.55.15832} {\bibfield  {journal}
  {\bibinfo  {journal} {Physical Review B}\ }\textbf {\bibinfo {volume} {55}},\
  \bibinfo {pages} {15832} (\bibinfo {year} {1997})}\BibitemShut {NoStop}%
\bibitem [{\citenamefont {Bid}\ \emph {et~al.}(2010)\citenamefont {Bid},
  \citenamefont {Ofek}, \citenamefont {Inoue}, \citenamefont {Heiblum},
  \citenamefont {Kane}, \citenamefont {Umansky},\ and\ \citenamefont
  {Mahalu}}]{Bid2010}%
  \BibitemOpen
  \bibfield  {author} {\bibinfo {author} {\bibfnamefont {A.}~\bibnamefont
  {Bid}}, \bibinfo {author} {\bibfnamefont {N.}~\bibnamefont {Ofek}}, \bibinfo
  {author} {\bibfnamefont {H.}~\bibnamefont {Inoue}}, \bibinfo {author}
  {\bibfnamefont {M.}~\bibnamefont {Heiblum}}, \bibinfo {author} {\bibfnamefont
  {C.~L.}\ \bibnamefont {Kane}}, \bibinfo {author} {\bibfnamefont
  {V.}~\bibnamefont {Umansky}}, \ and\ \bibinfo {author} {\bibfnamefont
  {D.}~\bibnamefont {Mahalu}},\ }\href {\doibase 10.1038/nature09277}
  {\bibfield  {journal} {\bibinfo  {journal} {Nature}\ }\textbf {\bibinfo
  {volume} {466}},\ \bibinfo {pages} {585} (\bibinfo {year}
  {2010})}\BibitemShut {NoStop}%
\bibitem [{\citenamefont {Dolev}\ \emph {et~al.}(2011)\citenamefont {Dolev},
  \citenamefont {Gross}, \citenamefont {Sabo}, \citenamefont {Gurman},
  \citenamefont {Heiblum}, \citenamefont {Umansky},\ and\ \citenamefont
  {Mahalu}}]{Dolev2011}%
  \BibitemOpen
  \bibfield  {author} {\bibinfo {author} {\bibfnamefont {M.}~\bibnamefont
  {Dolev}}, \bibinfo {author} {\bibfnamefont {Y.}~\bibnamefont {Gross}},
  \bibinfo {author} {\bibfnamefont {R.}~\bibnamefont {Sabo}}, \bibinfo {author}
  {\bibfnamefont {I.}~\bibnamefont {Gurman}}, \bibinfo {author} {\bibfnamefont
  {M.}~\bibnamefont {Heiblum}}, \bibinfo {author} {\bibfnamefont
  {V.}~\bibnamefont {Umansky}}, \ and\ \bibinfo {author} {\bibfnamefont
  {D.}~\bibnamefont {Mahalu}},\ }\href {\doibase
  10.1103/PhysRevLett.107.036805} {\bibfield  {journal} {\bibinfo  {journal}
  {Physical Review Letters}\ }\textbf {\bibinfo {volume} {107}},\ \bibinfo
  {pages} {036805} (\bibinfo {year} {2011})}\BibitemShut {NoStop}%
\bibitem [{\citenamefont {Gross}\ \emph {et~al.}(2012)\citenamefont {Gross},
  \citenamefont {Dolev}, \citenamefont {Heiblum}, \citenamefont {Umansky},\
  and\ \citenamefont {Mahalu}}]{Gross2012}%
  \BibitemOpen
  \bibfield  {author} {\bibinfo {author} {\bibfnamefont {Y.}~\bibnamefont
  {Gross}}, \bibinfo {author} {\bibfnamefont {M.}~\bibnamefont {Dolev}},
  \bibinfo {author} {\bibfnamefont {M.}~\bibnamefont {Heiblum}}, \bibinfo
  {author} {\bibfnamefont {V.}~\bibnamefont {Umansky}}, \ and\ \bibinfo
  {author} {\bibfnamefont {D.}~\bibnamefont {Mahalu}},\ }\href {\doibase
  10.1103/PhysRevLett.108.226801} {\bibfield  {journal} {\bibinfo  {journal}
  {Physical Review Letters}\ }\textbf {\bibinfo {volume} {108}},\ \bibinfo
  {pages} {226801} (\bibinfo {year} {2012})}\BibitemShut {NoStop}%
\bibitem [{\citenamefont {Inoue}\ \emph {et~al.}(2014)\citenamefont {Inoue},
  \citenamefont {Grivnin}, \citenamefont {Ronen}, \citenamefont {Heiblum},
  \citenamefont {Umansky},\ and\ \citenamefont {Mahalu}}]{Inoue2014}%
  \BibitemOpen
  \bibfield  {author} {\bibinfo {author} {\bibfnamefont {H.}~\bibnamefont
  {Inoue}}, \bibinfo {author} {\bibfnamefont {A.}~\bibnamefont {Grivnin}},
  \bibinfo {author} {\bibfnamefont {Y.}~\bibnamefont {Ronen}}, \bibinfo
  {author} {\bibfnamefont {M.}~\bibnamefont {Heiblum}}, \bibinfo {author}
  {\bibfnamefont {V.}~\bibnamefont {Umansky}}, \ and\ \bibinfo {author}
  {\bibfnamefont {D.}~\bibnamefont {Mahalu}},\ }\href {\doibase
  10.1038/ncomms5067} {\bibfield  {journal} {\bibinfo  {journal} {Nature
  Communications}\ }\textbf {\bibinfo {volume} {5}},\ \bibinfo {pages} {4067}
  (\bibinfo {year} {2014})}\BibitemShut {NoStop}%
\bibitem [{\citenamefont {Venkatachalam}\ \emph {et~al.}(2012)\citenamefont
  {Venkatachalam}, \citenamefont {Hart}, \citenamefont {Pfeiffer},
  \citenamefont {West},\ and\ \citenamefont {Yacoby}}]{Venkatachalam2012}%
  \BibitemOpen
  \bibfield  {author} {\bibinfo {author} {\bibfnamefont {V.}~\bibnamefont
  {Venkatachalam}}, \bibinfo {author} {\bibfnamefont {S.}~\bibnamefont {Hart}},
  \bibinfo {author} {\bibfnamefont {L.}~\bibnamefont {Pfeiffer}}, \bibinfo
  {author} {\bibfnamefont {K.}~\bibnamefont {West}}, \ and\ \bibinfo {author}
  {\bibfnamefont {A.}~\bibnamefont {Yacoby}},\ }\href {\doibase
  10.1038/nphys2384} {\bibfield  {journal} {\bibinfo  {journal} {Nature
  Physics}\ }\textbf {\bibinfo {volume} {8}},\ \bibinfo {pages} {676} (\bibinfo
  {year} {2012})}\BibitemShut {NoStop}%
\bibitem [{\citenamefont {Banerjee}\ \emph {et~al.}(2017)\citenamefont
  {Banerjee}, \citenamefont {Heiblum}, \citenamefont {Rosenblatt},
  \citenamefont {Oreg}, \citenamefont {Feldman}, \citenamefont {Stern},\ and\
  \citenamefont {Umansky}}]{Banerjee2017}%
  \BibitemOpen
  \bibfield  {author} {\bibinfo {author} {\bibfnamefont {M.}~\bibnamefont
  {Banerjee}}, \bibinfo {author} {\bibfnamefont {M.}~\bibnamefont {Heiblum}},
  \bibinfo {author} {\bibfnamefont {A.}~\bibnamefont {Rosenblatt}}, \bibinfo
  {author} {\bibfnamefont {Y.}~\bibnamefont {Oreg}}, \bibinfo {author}
  {\bibfnamefont {D.~E.}\ \bibnamefont {Feldman}}, \bibinfo {author}
  {\bibfnamefont {A.}~\bibnamefont {Stern}}, \ and\ \bibinfo {author}
  {\bibfnamefont {V.}~\bibnamefont {Umansky}},\ }\href {\doibase
  10.1038/nature22052} {\bibfield  {journal} {\bibinfo  {journal} {Nature}\
  }\textbf {\bibinfo {volume} {545}},\ \bibinfo {pages} {75} (\bibinfo {year}
  {2017})}\BibitemShut {NoStop}%
\bibitem [{\citenamefont {Banerjee}\ \emph {et~al.}(2018)\citenamefont
  {Banerjee}, \citenamefont {Heiblum}, \citenamefont {Umansky}, \citenamefont
  {Feldman}, \citenamefont {Oreg},\ and\ \citenamefont {Stern}}]{Banerjee2018}%
  \BibitemOpen
  \bibfield  {author} {\bibinfo {author} {\bibfnamefont {M.}~\bibnamefont
  {Banerjee}}, \bibinfo {author} {\bibfnamefont {M.}~\bibnamefont {Heiblum}},
  \bibinfo {author} {\bibfnamefont {V.}~\bibnamefont {Umansky}}, \bibinfo
  {author} {\bibfnamefont {D.~E.}\ \bibnamefont {Feldman}}, \bibinfo {author}
  {\bibfnamefont {Y.}~\bibnamefont {Oreg}}, \ and\ \bibinfo {author}
  {\bibfnamefont {A.}~\bibnamefont {Stern}},\ }\href {\doibase
  10.1038/s41586-018-0184-1} {\bibfield  {journal} {\bibinfo  {journal}
  {Nature}\ }\textbf {\bibinfo {volume} {559}},\ \bibinfo {pages} {205}
  (\bibinfo {year} {2018})}\BibitemShut {NoStop}%
\bibitem [{\citenamefont {Srivastav}\ \emph {et~al.}(2021)\citenamefont
  {Srivastav}, \citenamefont {Kumar}, \citenamefont {Sp{\aa}nsl{\"{a}}tt},
  \citenamefont {Watanabe}, \citenamefont {Taniguchi}, \citenamefont {Mirlin},
  \citenamefont {Gefen},\ and\ \citenamefont {Das}}]{Srivastav2021}%
  \BibitemOpen
  \bibfield  {author} {\bibinfo {author} {\bibfnamefont {S.~K.}\ \bibnamefont
  {Srivastav}}, \bibinfo {author} {\bibfnamefont {R.}~\bibnamefont {Kumar}},
  \bibinfo {author} {\bibfnamefont {C.}~\bibnamefont {Sp{\aa}nsl{\"{a}}tt}},
  \bibinfo {author} {\bibfnamefont {K.}~\bibnamefont {Watanabe}}, \bibinfo
  {author} {\bibfnamefont {T.}~\bibnamefont {Taniguchi}}, \bibinfo {author}
  {\bibfnamefont {A.~D.}\ \bibnamefont {Mirlin}}, \bibinfo {author}
  {\bibfnamefont {Y.}~\bibnamefont {Gefen}}, \ and\ \bibinfo {author}
  {\bibfnamefont {A.}~\bibnamefont {Das}},\ }\href {\doibase
  10.1103/PhysRevLett.126.216803} {\bibfield  {journal} {\bibinfo  {journal}
  {Physical Review Letters}\ }\textbf {\bibinfo {volume} {126}},\ \bibinfo
  {pages} {216803} (\bibinfo {year} {2021})}\BibitemShut {NoStop}%
\bibitem [{\citenamefont {Melcer}\ \emph {et~al.}(2022)\citenamefont {Melcer},
  \citenamefont {Dutta}, \citenamefont {Sp{\aa}nsl{\"{a}}tt}, \citenamefont
  {Park}, \citenamefont {Mirlin},\ and\ \citenamefont {Umansky}}]{Melcer2022}%
  \BibitemOpen
  \bibfield  {author} {\bibinfo {author} {\bibfnamefont {R.~A.}\ \bibnamefont
  {Melcer}}, \bibinfo {author} {\bibfnamefont {B.}~\bibnamefont {Dutta}},
  \bibinfo {author} {\bibfnamefont {C.}~\bibnamefont {Sp{\aa}nsl{\"{a}}tt}},
  \bibinfo {author} {\bibfnamefont {J.}~\bibnamefont {Park}}, \bibinfo {author}
  {\bibfnamefont {A.~D.}\ \bibnamefont {Mirlin}}, \ and\ \bibinfo {author}
  {\bibfnamefont {V.}~\bibnamefont {Umansky}},\ }\href {\doibase
  10.1038/s41467-022-28009-0} {\bibfield  {journal} {\bibinfo  {journal}
  {Nature Communications}\ }\textbf {\bibinfo {volume} {13}},\ \bibinfo {pages}
  {376} (\bibinfo {year} {2022})}\BibitemShut {NoStop}%
\bibitem [{\citenamefont {Protopopov}\ \emph {et~al.}(2017)\citenamefont
  {Protopopov}, \citenamefont {Gefen},\ and\ \citenamefont
  {Mirlin}}]{Protopopov2017}%
  \BibitemOpen
  \bibfield  {author} {\bibinfo {author} {\bibfnamefont {I.}~\bibnamefont
  {Protopopov}}, \bibinfo {author} {\bibfnamefont {Y.}~\bibnamefont {Gefen}}, \
  and\ \bibinfo {author} {\bibfnamefont {A.}~\bibnamefont {Mirlin}},\ }\href
  {\doibase 10.1016/J.AOP.2017.07.015} {\bibfield  {journal} {\bibinfo
  {journal} {Annals of Physics}\ }\textbf {\bibinfo {volume} {385}},\ \bibinfo
  {pages} {287} (\bibinfo {year} {2017})}\BibitemShut {NoStop}%
\bibitem [{\citenamefont {Nosiglia}\ \emph {et~al.}(2018)\citenamefont
  {Nosiglia}, \citenamefont {Park}, \citenamefont {Rosenow},\ and\
  \citenamefont {Gefen}}]{Nosiglia2018}%
  \BibitemOpen
  \bibfield  {author} {\bibinfo {author} {\bibfnamefont {C.}~\bibnamefont
  {Nosiglia}}, \bibinfo {author} {\bibfnamefont {J.}~\bibnamefont {Park}},
  \bibinfo {author} {\bibfnamefont {B.}~\bibnamefont {Rosenow}}, \ and\
  \bibinfo {author} {\bibfnamefont {Y.}~\bibnamefont {Gefen}},\ }\href
  {\doibase 10.1103/PhysRevB.98.115408} {\bibfield  {journal} {\bibinfo
  {journal} {Physical Review B}\ }\textbf {\bibinfo {volume} {98}},\ \bibinfo
  {pages} {115408} (\bibinfo {year} {2018})}\BibitemShut {NoStop}%
\bibitem [{\citenamefont {Ma}\ and\ \citenamefont {Feldman}(2019)}]{Ma2019}%
  \BibitemOpen
  \bibfield  {author} {\bibinfo {author} {\bibfnamefont {K.~K.~W.}\
  \bibnamefont {Ma}}\ and\ \bibinfo {author} {\bibfnamefont {D.~E.}\
  \bibnamefont {Feldman}},\ }\href {\doibase 10.1103/PhysRevB.99.085309}
  {\bibfield  {journal} {\bibinfo  {journal} {Physical Review B}\ }\textbf
  {\bibinfo {volume} {99}},\ \bibinfo {pages} {085309} (\bibinfo {year}
  {2019})}\BibitemShut {NoStop}%
\bibitem [{\citenamefont {Sp{\aa}nsl{\"{a}}tt}\ \emph
  {et~al.}(2019)\citenamefont {Sp{\aa}nsl{\"{a}}tt}, \citenamefont {Park},
  \citenamefont {Gefen},\ and\ \citenamefont {Mirlin}}]{Spanslatt2019}%
  \BibitemOpen
  \bibfield  {author} {\bibinfo {author} {\bibfnamefont {C.}~\bibnamefont
  {Sp{\aa}nsl{\"{a}}tt}}, \bibinfo {author} {\bibfnamefont {J.}~\bibnamefont
  {Park}}, \bibinfo {author} {\bibfnamefont {Y.}~\bibnamefont {Gefen}}, \ and\
  \bibinfo {author} {\bibfnamefont {A.~D.}\ \bibnamefont {Mirlin}},\ }\href
  {\doibase 10.1103/PhysRevLett.123.137701} {\bibfield  {journal} {\bibinfo
  {journal} {Physical Review Letters}\ }\textbf {\bibinfo {volume} {123}},\
  \bibinfo {pages} {137701} (\bibinfo {year} {2019})}\BibitemShut {NoStop}%
\bibitem [{\citenamefont {Aharon-Steinberg}\ \emph {et~al.}(2019)\citenamefont
  {Aharon-Steinberg}, \citenamefont {Oreg},\ and\ \citenamefont
  {Stern}}]{Aharon-Steinberg2019}%
  \BibitemOpen
  \bibfield  {author} {\bibinfo {author} {\bibfnamefont {A.}~\bibnamefont
  {Aharon-Steinberg}}, \bibinfo {author} {\bibfnamefont {Y.}~\bibnamefont
  {Oreg}}, \ and\ \bibinfo {author} {\bibfnamefont {A.}~\bibnamefont {Stern}},\
  }\href {\doibase 10.1103/PhysRevB.99.041302} {\bibfield  {journal} {\bibinfo
  {journal} {Physical Review B}\ }\textbf {\bibinfo {volume} {99}},\ \bibinfo
  {pages} {041302(R)} (\bibinfo {year} {2019})}\BibitemShut {NoStop}%
\bibitem [{\citenamefont {Ma}\ and\ \citenamefont {Feldman}(2020)}]{Ma2020}%
  \BibitemOpen
  \bibfield  {author} {\bibinfo {author} {\bibfnamefont {K.~K.~W.}\ \bibnamefont
  {Ma}}\ and\ \bibinfo {author} {\bibfnamefont {D.}~\bibnamefont {Feldman}},\
  }\href {\doibase 10.1103/PhysRevLett.125.016801} {\bibfield  {journal}
  {\bibinfo  {journal} {Physical Review Letters}\ }\textbf {\bibinfo {volume}
  {125}},\ \bibinfo {pages} {016801} (\bibinfo {year} {2020})}\BibitemShut
  {NoStop}%
\bibitem [{\citenamefont {Cohen}\ \emph {et~al.}(2019)\citenamefont {Cohen},
  \citenamefont {Ronen}, \citenamefont {Yang}, \citenamefont {Banitt},
  \citenamefont {Park}, \citenamefont {Heiblum}, \citenamefont {Mirlin},
  \citenamefont {Gefen},\ and\ \citenamefont {Umansky}}]{Cohen2019}%
  \BibitemOpen
  \bibfield  {author} {\bibinfo {author} {\bibfnamefont {Y.}~\bibnamefont
  {Cohen}}, \bibinfo {author} {\bibfnamefont {Y.}~\bibnamefont {Ronen}},
  \bibinfo {author} {\bibfnamefont {W.}~\bibnamefont {Yang}}, \bibinfo {author}
  {\bibfnamefont {D.}~\bibnamefont {Banitt}}, \bibinfo {author} {\bibfnamefont
  {J.}~\bibnamefont {Park}}, \bibinfo {author} {\bibfnamefont {M.}~\bibnamefont
  {Heiblum}}, \bibinfo {author} {\bibfnamefont {A.~D.}\ \bibnamefont {Mirlin}},
  \bibinfo {author} {\bibfnamefont {Y.}~\bibnamefont {Gefen}}, \ and\ \bibinfo
  {author} {\bibfnamefont {V.}~\bibnamefont {Umansky}},\ }\href {\doibase
  10.1038/s41467-019-09920-5} {\bibfield  {journal} {\bibinfo  {journal}
  {Nature Communications}\ }\textbf {\bibinfo {volume} {10}},\ \bibinfo {pages}
  {1920} (\bibinfo {year} {2019})}\BibitemShut {NoStop}%
\bibitem [{\citenamefont {Lafont}\ \emph {et~al.}(2019)\citenamefont {Lafont},
  \citenamefont {Rosenblatt}, \citenamefont {Heiblum},\ and\ \citenamefont
  {Umansky}}]{Lafont2019}%
  \BibitemOpen
  \bibfield  {author} {\bibinfo {author} {\bibfnamefont {F.}~\bibnamefont
  {Lafont}}, \bibinfo {author} {\bibfnamefont {A.}~\bibnamefont {Rosenblatt}},
  \bibinfo {author} {\bibfnamefont {M.}~\bibnamefont {Heiblum}}, \ and\
  \bibinfo {author} {\bibfnamefont {V.}~\bibnamefont {Umansky}},\ }\href
  {\doibase 10.1126/science.aar3766} {\bibfield  {journal} {\bibinfo  {journal}
  {Science}\ }\textbf {\bibinfo {volume} {363}},\ \bibinfo {pages} {54}
  (\bibinfo {year} {2019})}\BibitemShut {NoStop}%
\bibitem [{\citenamefont {Srivastav}\ \emph {et~al.}(2022)\citenamefont
  {Srivastav}, \citenamefont {Kumar}, \citenamefont {Sp{\aa}nsl{\"{a}}tt},
  \citenamefont {Watanabe}, \citenamefont {Taniguchi}, \citenamefont {Mirlin},
  \citenamefont {Gefen},\ and\ \citenamefont {Das}}]{Srivastav2022}%
  \BibitemOpen
  \bibfield  {author} {\bibinfo {author} {\bibfnamefont {S.~K.}\ \bibnamefont
  {Srivastav}}, \bibinfo {author} {\bibfnamefont {R.}~\bibnamefont {Kumar}},
  \bibinfo {author} {\bibfnamefont {C.}~\bibnamefont {Sp{\aa}nsl{\"{a}}tt}},
  \bibinfo {author} {\bibfnamefont {K.}~\bibnamefont {Watanabe}}, \bibinfo
  {author} {\bibfnamefont {T.}~\bibnamefont {Taniguchi}}, \bibinfo {author}
  {\bibfnamefont {A.~D.}\ \bibnamefont {Mirlin}}, \bibinfo {author}
  {\bibfnamefont {Y.}~\bibnamefont {Gefen}}, \ and\ \bibinfo {author}
  {\bibfnamefont {A.}~\bibnamefont {Das}},\ }\href
  {http://arxiv.org/abs/2202.00490} {\  (\bibinfo {year} {2022})},\ \Eprint
  {http://arxiv.org/abs/2202.00490} {arXiv:2202.00490} \BibitemShut {NoStop}%
\bibitem [{\citenamefont {Jezouin}\ \emph {et~al.}(2013)\citenamefont
  {Jezouin}, \citenamefont {Parmentier}, \citenamefont {Anthore}, \citenamefont
  {Gennser}, \citenamefont {Cavanna}, \citenamefont {Jin},\ and\ \citenamefont
  {Pierre}}]{Jezouin2013a}%
  \BibitemOpen
  \bibfield  {author} {\bibinfo {author} {\bibfnamefont {S.}~\bibnamefont
  {Jezouin}}, \bibinfo {author} {\bibfnamefont {F.~D.}\ \bibnamefont
  {Parmentier}}, \bibinfo {author} {\bibfnamefont {A.}~\bibnamefont {Anthore}},
  \bibinfo {author} {\bibfnamefont {U.}~\bibnamefont {Gennser}}, \bibinfo
  {author} {\bibfnamefont {A.}~\bibnamefont {Cavanna}}, \bibinfo {author}
  {\bibfnamefont {Y.}~\bibnamefont {Jin}}, \ and\ \bibinfo {author}
  {\bibfnamefont {F.}~\bibnamefont {Pierre}},\ }\href {\doibase
  10.1126/science.1241912} {\bibfield  {journal} {\bibinfo  {journal} {Science
  (New York, N.Y.)}\ }\textbf {\bibinfo {volume} {342}},\ \bibinfo {pages}
  {601} (\bibinfo {year} {2013})}\BibitemShut {NoStop}%
\bibitem [{\citenamefont {Srivastav}\ \emph {et~al.}(2019)\citenamefont
  {Srivastav}, \citenamefont {Sahu}, \citenamefont {Watanabe}, \citenamefont
  {Taniguchi}, \citenamefont {Banerjee},\ and\ \citenamefont
  {Das}}]{Srivastav2019}%
  \BibitemOpen
  \bibfield  {author} {\bibinfo {author} {\bibfnamefont {S.~K.}\ \bibnamefont
  {Srivastav}}, \bibinfo {author} {\bibfnamefont {M.~R.}\ \bibnamefont {Sahu}},
  \bibinfo {author} {\bibfnamefont {K.}~\bibnamefont {Watanabe}}, \bibinfo
  {author} {\bibfnamefont {T.}~\bibnamefont {Taniguchi}}, \bibinfo {author}
  {\bibfnamefont {S.}~\bibnamefont {Banerjee}}, \ and\ \bibinfo {author}
  {\bibfnamefont {A.}~\bibnamefont {Das}},\ }\href {\doibase
  10.1126/sciadv.aaw5798} {\bibfield  {journal} {\bibinfo  {journal} {Science
  Advances}\ }\textbf {\bibinfo {volume} {5}},\ \bibinfo {pages} {eaaw5798}
  (\bibinfo {year} {2019})}\BibitemShut {NoStop}%
\bibitem [{SM()}]{SM}%
  \BibitemOpen
  \href@noop {} {\ }\bibinfo {note} {See Supplemental Material at [url] for detailed descriptions of the setup, calibration procedures and measurements, as well as additional analysis and discussion, which includes Refs. [34-37].}\BibitemShut
  {Stop}%
\bibitem [{\citenamefont {Pendry}(1983)}]{Pendry1983}%
  \BibitemOpen
  \bibfield  {author} {\bibinfo {author} {\bibfnamefont {J.~B.}\ \bibnamefont
  {Pendry}},\ }\href {\doibase 10.1088/0305-4470/16/10/012} {\bibfield
  {journal} {\bibinfo  {journal} {Journal of Physics A: Mathematical and
  General}\ }\textbf {\bibinfo {volume} {16}},\ \bibinfo {pages} {2161}
  (\bibinfo {year} {1983})}\BibitemShut {NoStop}%
\bibitem [{\citenamefont {Rego}\ and\ \citenamefont
  {Kirczenow}(1999)}]{Rego1999}%
  \BibitemOpen
  \bibfield  {author} {\bibinfo {author} {\bibfnamefont {L.~G.~C.}\
  \bibnamefont {Rego}}\ and\ \bibinfo {author} {\bibfnamefont {G.}~\bibnamefont
  {Kirczenow}},\ }\href {\doibase 10.1103/PhysRevB.59.13080} {\bibfield
  {journal} {\bibinfo  {journal} {Physical Review B}\ }\textbf {\bibinfo
  {volume} {59}},\ \bibinfo {pages} {13080} (\bibinfo {year}
  {1999})}\BibitemShut {NoStop}%
\bibitem [{\citenamefont {Sivre}\ \emph {et~al.}(2019)\citenamefont {Sivre},
  \citenamefont {Duprez}, \citenamefont {Anthore}, \citenamefont {Aassime},
  \citenamefont {Parmentier}, \citenamefont {Cavanna}, \citenamefont {Ouerghi},
  \citenamefont {Gennser},\ and\ \citenamefont {Pierre}}]{Sivre2019}%
  \BibitemOpen
  \bibfield  {author} {\bibinfo {author} {\bibfnamefont {E.}~\bibnamefont
  {Sivre}}, \bibinfo {author} {\bibfnamefont {H.}~\bibnamefont {Duprez}},
  \bibinfo {author} {\bibfnamefont {A.}~\bibnamefont {Anthore}}, \bibinfo
  {author} {\bibfnamefont {A.}~\bibnamefont {Aassime}}, \bibinfo {author}
  {\bibfnamefont {F.~D.}\ \bibnamefont {Parmentier}}, \bibinfo {author}
  {\bibfnamefont {A.}~\bibnamefont {Cavanna}}, \bibinfo {author} {\bibfnamefont
  {A.}~\bibnamefont {Ouerghi}}, \bibinfo {author} {\bibfnamefont
  {U.}~\bibnamefont {Gennser}}, \ and\ \bibinfo {author} {\bibfnamefont
  {F.}~\bibnamefont {Pierre}},\ }\href {\doibase 10.1038/s41467-019-13566-8}
  {\bibfield  {journal} {\bibinfo  {journal} {Nature Communications}\ }\textbf
  {\bibinfo {volume} {10}},\ \bibinfo {pages} {5638} (\bibinfo {year}
  {2019})}\BibitemShut {NoStop}%
\bibitem [{\citenamefont {Iftikhar}\ \emph {et~al.}(2016)\citenamefont
  {Iftikhar}, \citenamefont {Anthore}, \citenamefont {Jezouin}, \citenamefont
  {Parmentier}, \citenamefont {Jin}, \citenamefont {Cavanna}, \citenamefont
  {Ouerghi}, \citenamefont {Gennser},\ and\ \citenamefont
  {Pierre}}]{Iftikhar2016}%
  \BibitemOpen
  \bibfield  {author} {\bibinfo {author} {\bibfnamefont {Z.}~\bibnamefont
  {Iftikhar}}, \bibinfo {author} {\bibfnamefont {A.}~\bibnamefont {Anthore}},
  \bibinfo {author} {\bibfnamefont {S.}~\bibnamefont {Jezouin}}, \bibinfo
  {author} {\bibfnamefont {F.}~\bibnamefont {Parmentier}}, \bibinfo {author}
  {\bibfnamefont {Y.}~\bibnamefont {Jin}}, \bibinfo {author} {\bibfnamefont
  {A.}~\bibnamefont {Cavanna}}, \bibinfo {author} {\bibfnamefont
  {A.}~\bibnamefont {Ouerghi}}, \bibinfo {author} {\bibfnamefont
  {U.}~\bibnamefont {Gennser}}, \ and\ \bibinfo {author} {\bibfnamefont
  {F.}~\bibnamefont {Pierre}},\ }\href {\doibase 10.1038/ncomms12908}
  {\bibfield  {journal} {\bibinfo  {journal} {Nature Communications}\ }\textbf
  {\bibinfo {volume} {7}} (\bibinfo {year} {2016}),\
  10.1038/ncomms12908}\BibitemShut {NoStop}%
\bibitem [{\citenamefont {Sivre}\ \emph {et~al.}(2018)\citenamefont {Sivre},
  \citenamefont {Anthore}, \citenamefont {Parmentier}, \citenamefont {Cavanna},
  \citenamefont {Gennser}, \citenamefont {Ouerghi}, \citenamefont {Jin},\ and\
  \citenamefont {Pierre}}]{Sivre2018}%
  \BibitemOpen
  \bibfield  {author} {\bibinfo {author} {\bibfnamefont {E.}~\bibnamefont
  {Sivre}}, \bibinfo {author} {\bibfnamefont {A.}~\bibnamefont {Anthore}},
  \bibinfo {author} {\bibfnamefont {F.~D.}\ \bibnamefont {Parmentier}},
  \bibinfo {author} {\bibfnamefont {A.}~\bibnamefont {Cavanna}}, \bibinfo
  {author} {\bibfnamefont {U.}~\bibnamefont {Gennser}}, \bibinfo {author}
  {\bibfnamefont {A.}~\bibnamefont {Ouerghi}}, \bibinfo {author} {\bibfnamefont
  {Y.}~\bibnamefont {Jin}}, \ and\ \bibinfo {author} {\bibfnamefont
  {F.}~\bibnamefont {Pierre}},\ }\href {\doibase 10.1038/nphys4280} {\bibfield
  {journal} {\bibinfo  {journal} {Nature Physics}\ }\textbf {\bibinfo {volume}
  {14}},\ \bibinfo {pages} {145} (\bibinfo {year} {2018})}\BibitemShut {NoStop}%
\bibitem [{\citenamefont {Amet}\ \emph {et~al.}(2014)\citenamefont {Amet},
  \citenamefont {Williams}, \citenamefont {Watanabe}, \citenamefont
  {Taniguchi},\ and\ \citenamefont {Goldhaber-Gordon}}]{Amet2014}%
  \BibitemOpen
  \bibfield  {author} {\bibinfo {author} {\bibfnamefont {F.}~\bibnamefont
  {Amet}}, \bibinfo {author} {\bibfnamefont {J.R.~\bibnamefont {Williams}},
  \bibinfo {author} {\bibfnamefont {K.}~\bibnamefont {Watanabe}}, \bibinfo
  {author} {\bibfnamefont {T.}~\bibnamefont {Taniguchi}}, \ and\ \bibinfo
  {author} {\bibfnamefont {D.}~\bibnamefont {Goldhaber-Gordon}},\ }}\href
  {\doibase 10.1103/PhysRevLett.112.196601} {\bibfield  {journal} {\bibinfo
  {journal} {Physical Review Letters}\ }\textbf {\bibinfo {volume} {112}},\
  \bibinfo {pages} {196601} (\bibinfo {year} {2014})}\BibitemShut {NoStop}%
\bibitem [{\citenamefont {Wei}\ \emph {et~al.}(2017)\citenamefont {Wei},
  \citenamefont {van~der Sar}, \citenamefont {Sanchez-Yamagishi}, \citenamefont
  {Watanabe}, \citenamefont {Taniguchi}, \citenamefont {Jarillo-Herrero},
  \citenamefont {Halperin},\ and\ \citenamefont {Yacoby}}]{Wei2017}%
  \BibitemOpen
  \bibfield  {author} {\bibinfo {author} {\bibfnamefont {D.~S.}\ \bibnamefont
  {Wei}}, \bibinfo {author} {\bibfnamefont {T.}~\bibnamefont {van~der Sar}},
  \bibinfo {author} {\bibfnamefont {J.~D.}\ \bibnamefont {Sanchez-Yamagishi}},
  \bibinfo {author} {\bibfnamefont {K.}~\bibnamefont {Watanabe}}, \bibinfo
  {author} {\bibfnamefont {T.}~\bibnamefont {Taniguchi}}, \bibinfo {author}
  {\bibfnamefont {P.}~\bibnamefont {Jarillo-Herrero}}, \bibinfo {author}
  {\bibfnamefont {B.~I.}\ \bibnamefont {Halperin}}, \ and\ \bibinfo {author}
  {\bibfnamefont {A.}~\bibnamefont {Yacoby}},\ }\href {\doibase
  10.1126/sciadv.1700600} {\bibfield  {journal} {\bibinfo  {journal} {Science
  Advances}\ }\textbf {\bibinfo {volume} {3}},\ \bibinfo {pages} {e1700600}
  (\bibinfo {year} {2017})}\BibitemShut {NoStop}%
\bibitem [{\citenamefont {Kim}\ \emph {et~al.}(2009)\citenamefont {Kim},
  \citenamefont {Nah}, \citenamefont {Jo}, \citenamefont {Shahrjerdi},
  \citenamefont {Colombo}, \citenamefont {Yao}, \citenamefont {Tutuc},\ and\
  \citenamefont {Banerjee}}]{Kim2009}%
  \BibitemOpen
  \bibfield  {author} {\bibinfo {author} {\bibfnamefont {S.}~\bibnamefont
  {Kim}}, \bibinfo {author} {\bibfnamefont {J.}~\bibnamefont {Nah}}, \bibinfo
  {author} {\bibfnamefont {I.}~\bibnamefont {Jo}}, \bibinfo {author}
  {\bibfnamefont {D.}~\bibnamefont {Shahrjerdi}}, \bibinfo {author}
  {\bibfnamefont {L.}~\bibnamefont {Colombo}}, \bibinfo {author} {\bibfnamefont
  {Z.}~\bibnamefont {Yao}}, \bibinfo {author} {\bibfnamefont {E.}~\bibnamefont
  {Tutuc}}, \ and\ \bibinfo {author} {\bibfnamefont {S.~K.}\ \bibnamefont
  {Banerjee}},\ }\href {\doibase 10.1063/1.3077021} {\bibfield  {journal}
  {\bibinfo  {journal} {Appl. Phys. Lett.}\ }\textbf {\bibinfo {volume} {94}},\
  \bibinfo {pages} {062107} (\bibinfo {year} {2009})}\BibitemShut {NoStop}%
\bibitem [{\citenamefont {Ribeiro-Palau}\ \emph {et~al.}(2019)\citenamefont
  {Ribeiro-Palau}, \citenamefont {Chen}, \citenamefont {Zeng}, \citenamefont
  {Watanabe}, \citenamefont {Taniguchi}, \citenamefont {Hone},\ and\
  \citenamefont {Dean}}]{Ribeiro-Palau2019}%
  \BibitemOpen
  \bibfield  {author} {\bibinfo {author} {\bibfnamefont {R.}~\bibnamefont
  {Ribeiro-Palau}}, \bibinfo {author} {\bibfnamefont {S.}~\bibnamefont {Chen}},
  \bibinfo {author} {\bibfnamefont {Y.}~\bibnamefont {Zeng}}, \bibinfo {author}
  {\bibfnamefont {K.}~\bibnamefont {Watanabe}}, \bibinfo {author}
  {\bibfnamefont {T.}~\bibnamefont {Taniguchi}}, \bibinfo {author}
  {\bibfnamefont {J.}~\bibnamefont {Hone}}, \ and\ \bibinfo {author}
  {\bibfnamefont {C.~R.}\ \bibnamefont {Dean}},\ }\href {\doibase
  10.1021/acs.nanolett.9b00351} {\bibfield  {journal} {\bibinfo  {journal}
  {Nano Lett.}\ }\textbf {\bibinfo {volume} {19}},\ \bibinfo {pages} {2583}
  (\bibinfo {year} {2019})}\BibitemShut {NoStop}%
\bibitem [{\citenamefont {Fl{\ifmmode\acute{o}\else\'{o}\fi}r}\ \emph
  {et~al.}(2022)\citenamefont {Fl{\ifmmode\acute{o}\else\'{o}\fi}r},
  \citenamefont {Lacerda-Santos}, \citenamefont {Fleury}, \citenamefont
  {Roulleau},\ and\ \citenamefont {Waintal}}]{Flor2022}%
  \BibitemOpen
  \bibfield  {author} {\bibinfo {author} {\bibfnamefont {I.~M.}\ \bibnamefont
  {Fl{\ifmmode\acute{o}\else\'{o}\fi}r}}, \bibinfo {author} {\bibfnamefont
  {A.}~\bibnamefont {Lacerda-Santos}}, \bibinfo {author} {\bibfnamefont
  {G.}~\bibnamefont {Fleury}}, \bibinfo {author} {\bibfnamefont
  {P.}~\bibnamefont {Roulleau}}, \ and\ \bibinfo {author} {\bibfnamefont
  {X.}~\bibnamefont {Waintal}},\ }\href {\doibase 10.1103/PhysRevB.105.L241409}
  {\bibfield  {journal} {\bibinfo  {journal} {Phys. Rev. B}\ }\textbf {\bibinfo
  {volume} {105}},\ \bibinfo {pages} {L241409} (\bibinfo {year}
  {2022})}\BibitemShut {NoStop}%
  \bibitem [{\citenamefont {Wei}\ \emph {et~al.}(2018)\citenamefont {Wei},
  \citenamefont {van~der Sar}, \citenamefont {Lee}, \citenamefont {Watanabe},
  \citenamefont {Taniguchi}, \citenamefont {Halperin},\ and\ \citenamefont
  {Yacoby}}]{Wei2018}%
  \BibitemOpen
  \bibfield  {author} {\bibinfo {author} {\bibfnamefont {D.~S.}\ \bibnamefont
  {Wei}}, \bibinfo {author} {\bibfnamefont {T.}~\bibnamefont {van~der Sar}},
  \bibinfo {author} {\bibfnamefont {S.~H.}\ \bibnamefont {Lee}}, \bibinfo
  {author} {\bibfnamefont {K.}~\bibnamefont {Watanabe}}, \bibinfo {author}
  {\bibfnamefont {T.}~\bibnamefont {Taniguchi}}, \bibinfo {author}
  {\bibfnamefont {B.~I.}\ \bibnamefont {Halperin}}, \ and\ \bibinfo {author}
  {\bibfnamefont {A.}~\bibnamefont {Yacoby}},\ }\href {\doibase
  10.1126/science.aar4061} {\bibfield  {journal} {\bibinfo  {journal}
  {Science}\ }\textbf {\bibinfo {volume} {362}},\ \bibinfo {pages} {229}
  (\bibinfo {year} {2018})}\BibitemShut {NoStop}%
\end{thebibliography}
\end{document}

\title{Supplemental Material for ''Heat equilibration of integer and fractional quantum Hall edge modes in graphene''}

\author{G. Le Breton}
\affiliation{Universit\'e Paris-Saclay, CEA, CNRS, SPEC, 91191 Gif-sur-Yvette cedex, France
}
\author{R. Delagrange}
\affiliation{Universit\'e Paris-Saclay, CEA, CNRS, SPEC, 91191 Gif-sur-Yvette cedex, France
}
\author{Y. Hong}
\affiliation{Universit\'e Paris-Saclay, CNRS, Centre de Nanosciences et de Nanotechnologies (C2N), 91120 Palaiseau, France
}
\author{M. Garg}
\affiliation{Universit\'e Paris-Saclay, CEA, CNRS, SPEC, 91191 Gif-sur-Yvette cedex, France
}
\author{K. Watanabe}
\affiliation{National Institute for Materials Science, Tsukuba, Japan
}
\author{T. Taniguchi}
\affiliation{National Institute for Materials Science, Tsukuba, Japan
}
\author{R. Ribeiro-Palau}
\affiliation{Universit\'e Paris-Saclay, CNRS, Centre de Nanosciences et de Nanotechnologies (C2N), 91120 Palaiseau, France
}
\author{P. Roulleau}
\affiliation{Universit\'e Paris-Saclay, CEA, CNRS, SPEC, 91191 Gif-sur-Yvette cedex, France
}
\author{P. Roche}
\affiliation{Universit\'e Paris-Saclay, CEA, CNRS, SPEC, 91191 Gif-sur-Yvette cedex, France
}
\author{F.D. Parmentier}
\affiliation{Universit\'e Paris-Saclay, CEA, CNRS, SPEC, 91191 Gif-sur-Yvette cedex, France
}

\date{\today}

%\begin{abstract}

%ABSTRACT
%\end{abstract}
{
\let\clearpage\relax
\maketitle
}

\section{Electronic heat balance calculations}

To obtain Eq. 1 in the main text, we first write the heat flow through a single integer quantum Hall edge channel stemming from an electron reservoir labelled $\alpha$ with chemical potential $\mu+eV_\alpha$ ($\mu$ is the global chemical potential of the sample in absence of dc bias) and temperature $T_\alpha$:

\begin{equation}
        J_\mathrm{out}^\alpha=\frac{1}{h}\int d\epsilon (\epsilon-\mu) \left[ f_\alpha (\epsilon) -\theta(\mu- \epsilon) \right]=\frac{\pi^2\kB^2T_\alpha^2}{6h}+\frac{1}{h}\frac{(eV_\alpha)^2}{2}=\frac{\kappa_0}{2}T_\alpha^2+\frac{G_0}{2}V_\alpha^2,
    \label{eq:heatflow}
\end{equation}

where $f_\alpha (\epsilon)$ is the Fermi function in the reservoir $\alpha$, and $\theta(\epsilon)$ is Heaviside's step function. This formula can be used to obtain the heat balance at integer filling factor $\nu$ in the central metallic island: $2\sum_1^\nu J_\mathrm{out}^c=\sum_1^\nu J_\mathrm{out}^I+\sum_1^\nu J_\mathrm{out}^G$, where $J_\mathrm{out}^I$ (resp. $J_\mathrm{out}^G$) is the heat flow carried by a single edge channel leaving the current feed contact upstream of the metallic island (resp. the upstream grounded contact on the other side of the metallic island), with temperature $T_0$ and chemical potential $\mu+\Idc/(\nu G_0)$ (resp. $\mu$). Recalling that $V_c=\Idc/(2 \nu G_0)$, this yields:

\begin{equation}
       2\sum_1^\nu\frac{\kappa_0}{2}T_c^2+2\sum_1^\nu \frac{G_0}{2} \left(\frac{\Idc}{2\nu G_0}\right)^2 =2\sum_1^\nu\frac{\kappa_0}{2}T_0^2+\sum_1^\nu \frac{G_0}{2} \left(\frac{\Idc}{\nu G_0}\right)^2.
    \label{eq:heatflow2}
\end{equation}

Grouping the temperature and dc current-dependent terms on either side of the equation gives:

\begin{equation}
        \sum_1^\nu \frac{\Idc^2}{4\nu^2 G_0}  =2\sum_1^\nu\frac{\kappa_0}{2}(T_c^2-T_0^2),
    \label{eq:heatflow3}
\end{equation}
which, in the case of integer filling factor where $\nu=N$, yields main text Eq. 1:

\begin{equation}
\frac{1}{4\nu G_0}\Idc^2 = 2 N \frac{\kappa_0}{2}(\Tc^2-T_0^2).
\label{eq:heatbalance}
\end{equation}

This equation can be generalized to the fractional case by singling out the contributions of the fractional charged and neutral channels.

\section{Sample fabrication and characterization}

The sample was made of a van der Waals heterostructure (36 nm h-BN / monolayer graphene / 25 nm h-BN / few nm graphite) assembled from top to bottom with a PDMS stamp covered by a PPC film. Metallic contacts, including the central island and the graphite gate contact, where defined using electron beam lithography. The top h-BN and about 10 nm of the bottom h-BN were etched using CHF3/O2 reactive ion etching to expose the graphene edge. Metallic electrodes (5 nm Ti / 100 nm Au) were then deposited in the etched trenches so as to connect the graphene. A second electron beam lithography was then realized to define the sample's edges using CHF3/O2 reactive ion etching.
The room temperature mobility of the sample was extracted from 4 points measurements for both sides of the sample, yielding $\mu=1.7\times10^5~$cm$^2/$Vs for the T-side, and $\mu=3.8\times10^5~$cm$^2/$Vs for the R-side. A small gate leakage was present above $\Vg>1~$V. 
After cooldown 1, the sample was rapidly warmed up to room temperature, leading to hoarfrost. It was then heated up to $150~^\circ$C for about 30 minutes, pumped to secondary vacuum ($\sim10^{-4}$~mbar) in the sample loadlock of our dilution refrigerator, then cooled down.

\begin{figure*}[h!]
\centering
\includegraphics[width=0.75\textwidth]{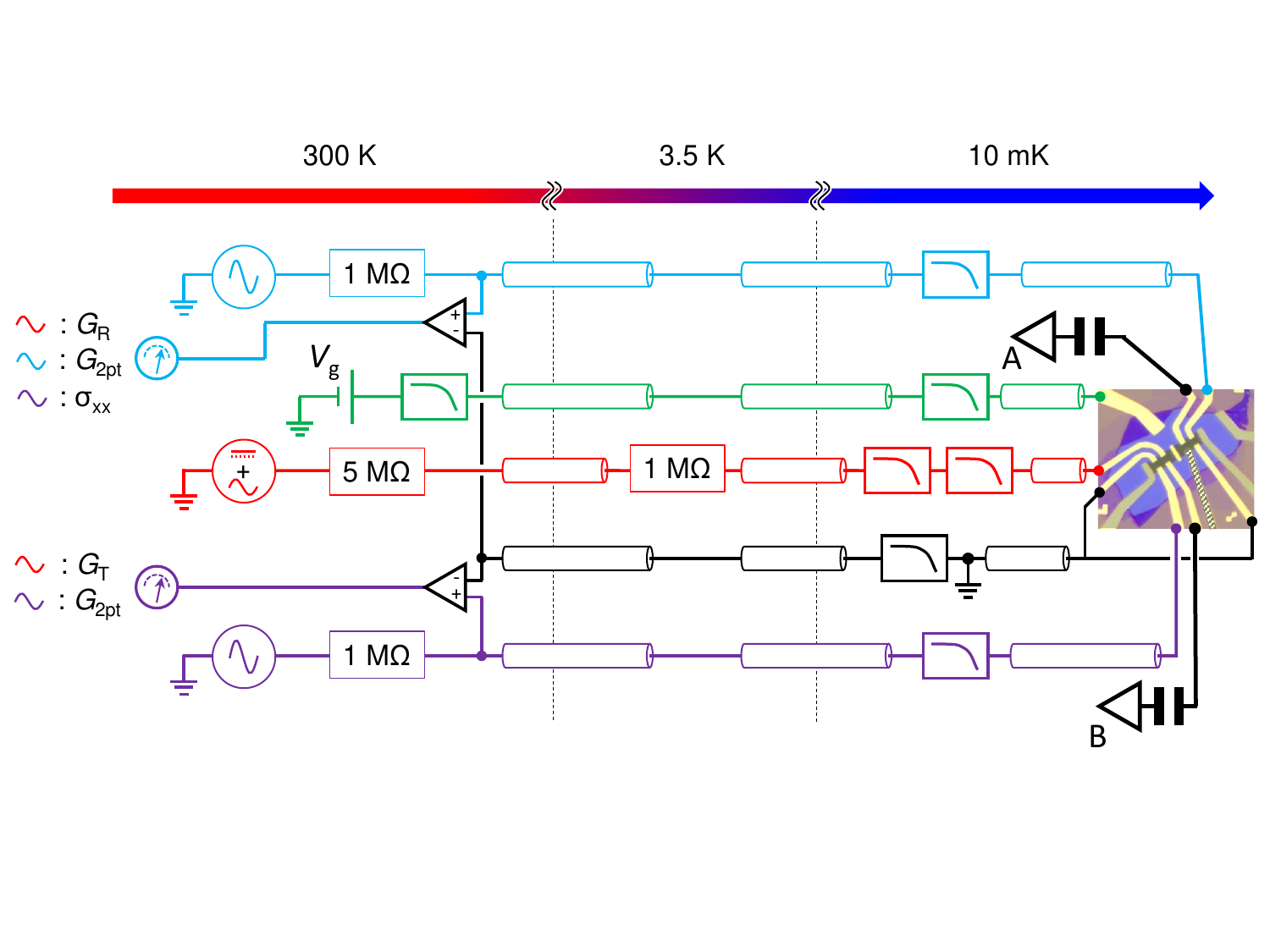}
\caption{\label{figsup-conductances} Layout of the wiring for the conductance measurements. Lines are color-coded (blue: T-side conductance; green: back gate; red: dc current feed; purple: R-side conductance; black: cold ground).}
\end{figure*}

\section{Conductance measurements}
\subsection{Setup}
A detailed description of the conductance measurements is shown in Fig.~\ref{figsup-conductances}. The measurements were performed using lock-in techniques at low frequency, below 10~Hz. All lines, including current feed (red in Fig.~\ref{figsup-conductances}) and back gate (green in Fig.~\ref{figsup-conductances}) are heavily filtered at the mixing chamber stage of our dilution refigerator using cascaded \textit{RC} filters. The effect of those filters (both in terms of series resistance and capacitive cutoff) are taken into account in our data. All measurements are performed using differential amplifiers (CELIANS EPC-1B) referenced to the cold ground (black in Fig.~\ref{figsup-conductances}) The latter is directly connected (both electrically and thermally) to the mixing chamber stage. The current feed line includes a $1~$M$\Omega$ series bias resistor thermally anchored to the $3.5~$K stage of our dilution refrigerator.

\subsection{Zero magnetic field measurements}

\begin{figure}[ht]
\centering
\includegraphics[width=0.3\textwidth]{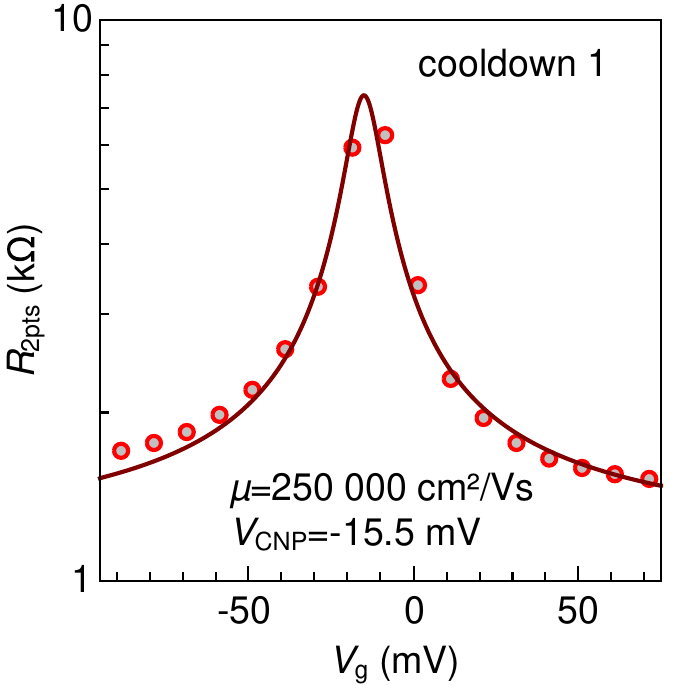}
\caption{\label{figsup-CNPcd1} Measurement of the 2-point resistance on the T side of the sample as a function of $\Vg$, at $B=0~$T and $T=12~$mK. Circles: data, line: fit.}
\end{figure}

Fig.~\ref{figsup-CNPcd1} shows the measurement of the sample resistance versus gate voltage, at zero magnetic field and low temperature, for cooldown 1. The charge neutrality point (CNP) is well fitted by the standard equation~\cite{Kim2009}, yielding a mobility of about $250000~$cm$^2/$Vs (this value is slightly different from the room temperature ones mentioned above, mostly because the two point configuration used here does not allow for a precise estimation of the number of squares used in the formula). The $\Vg$ position of the CNP is $-15.5~$mV. We were not able to perform this measurement for cooldown 2; nonetheless, we present below results of a third cooldown with conductance features similar to that of cooldown 2. 

\subsection{The $\nu=8/3$ plateau at cooldown 2}

\begin{figure}[ht]
\centering
\includegraphics[width=0.25\textwidth]{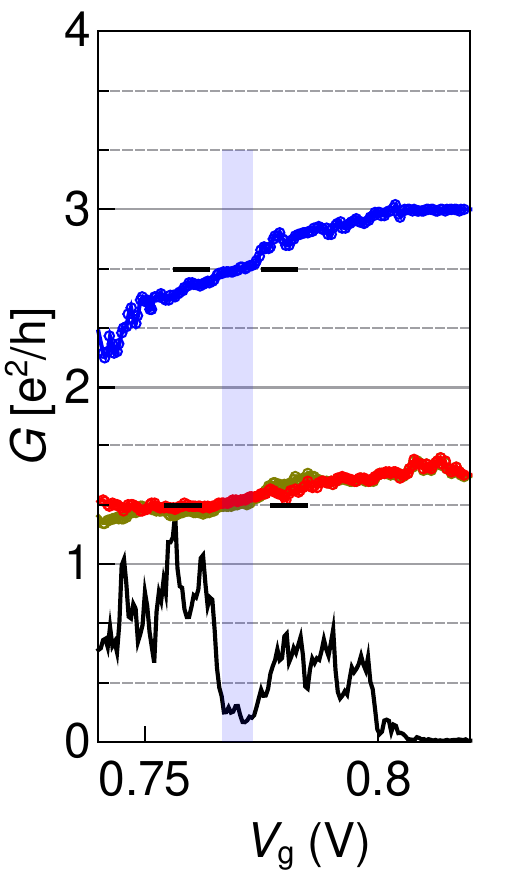}
\caption{\label{figsup-plateauzoom} Zoom on the $nu=8/3$ plateau on the conductance data at cooldown 2 shown in the main text. Blue: 2-point conductance $G_{2\mathrm{pt}}$ measured on the reflected side). Dark grey: longitudinal-like conductivity $\sigmaxx$. Orange/red: transmitted/reflected transconductance $G_\mathrm{T/R}$.}
\end{figure}

Figure \ref{figsup-plateauzoom} shows a more precise zoom on the $\nu=8/3$ plateau measured during cooldown 2. In particular, it shows that the criteria for chirality and current reditribution are well enforced even though the longitudinal-like conductivity is not exaclty zero.

\subsection{Gate voltage position of the plateaus and increase of the intrinsic doping}

\begin{figure}[ht]
\centering
\includegraphics[width=0.3\textwidth]{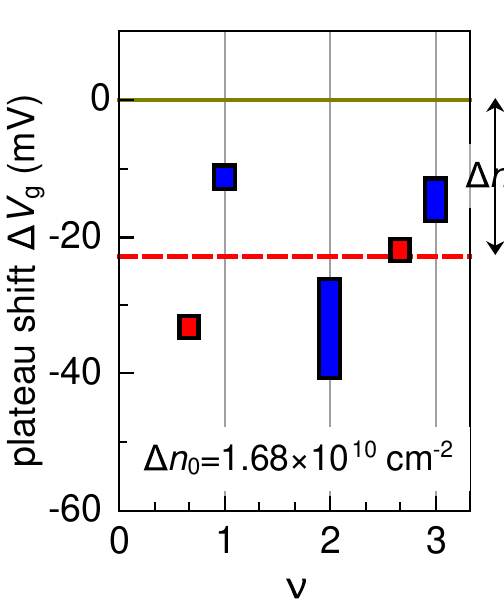}
\caption{\label{figsup-plateaushift} Gate voltage shift of the plateaus center at cooldown 2 with respect to cooldown 1. Fractional states are shown in red, integer in blue. The dashed line corresponds to the average shift, yielding the density increase $\Delta n_0$.}
\end{figure}

We infer the presence of increased intrinsic doping in the second cooldown by comparing the gate voltage positions of the quantum Hall plateaus for both cooldown, at the same value of the magnetic field $B=7~$T. In absence of zero magnetic field data for cooldown 2 (we were unfortunately forced to warm up the system before being able to perform these additional measurements), we estimate the intrinsic electron density increase $\Delta n_0$ from the shift of the plateaus center in cooldown 2 with respect to cooldown 1. The plateau center is obtained either from the 2 point conductances, or from the longitudinal conductance, yielding a typical uncertainty on this estimation. The shifts for filling factors $\nu=2/3$, 1, 2, $8/3$, and 3 are shown in Fig.~\ref{figsup-plateaushift}. All filling factors show a negative $\Vg$ shift, corresponding to a net increase of the intrinsic electron density. The value of this increase $\Delta n_0\approx 1.68\times 10^10~$cm$^{-2}$ is estimated from the average $\Delta \Vg\approx23~$mV shift, knowing the capacitive coupling between the graphite back gate and the graphene flake from the h-BN thickness. We argue that this increase is important enough to affect the edge electrostatics and strengthen the coupling between the integer and fractional edge channels at $\nu=8/3$. In a similar fashion, the slight increase in the width of the integer plateaus indicate an increase of the disorder, which remains small enough to allow us observing fractional quantum Hall states.

\subsection{Conductance versus dc current}

\begin{figure}[ht]
\centering
\includegraphics[width=0.5\textwidth]{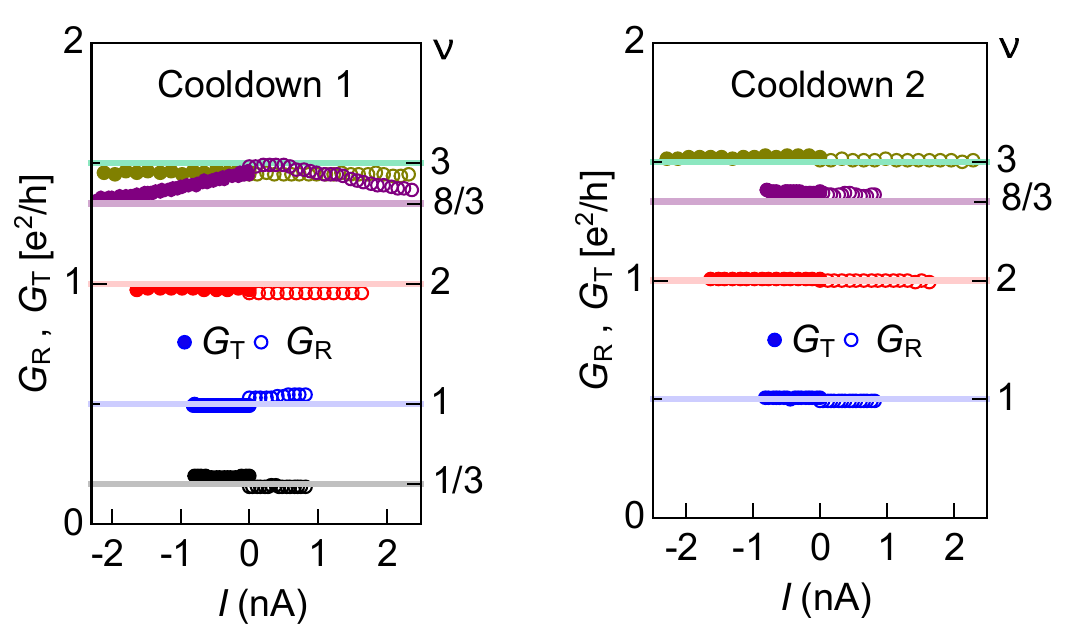}
\caption{\label{figsup-GvsI} Transmitted ($G_\mathrm{T}$) and reflected ($G_\mathrm{R}$) transconductances as a function of $\Idc$, for $\nu=1/3$ (black), $\nu=1$ (blue), $\nu=2$ (red), $\nu=8/3$ (purple), and $\nu=3$ (dark yellow). Measured $G_\mathrm{T}$ (full circles) are shown for negative $\Idc$, and $G_\mathrm{R}$ (open circles) for positive $\Idc$. Note that both $G_\mathrm{T}$ and $G_\mathrm{R}$ are essentially symmetric with the sign of $\Idc$. The lines are the expected values for $G_\mathrm{R,T}=G_0 \nu/2$.}
\end{figure}

We measured all conductances simultaneously to the noise measurement, allowing to check that the reflected and transmitted transconductance remains reasonably constant as a function of the applied $\Idc$. This is shown in Fig.~\ref{figsup-GvsI}, for cooldown 1 (left) and cooldown 2 (right). Except for $\nu=8/3$ in cooldown 2, which displays a variation of about $10~\%$ between $\Idc=\pm 2~$nA, all conductances are constant and very close to their expected values. Note that for this measurement, the ac excitation current used for the lock-in measurement was about $0.1~$nA.

\section{Noise measurements}

\begin{figure}[h!]
    \centering
    \includegraphics[width=0.7\textwidth]{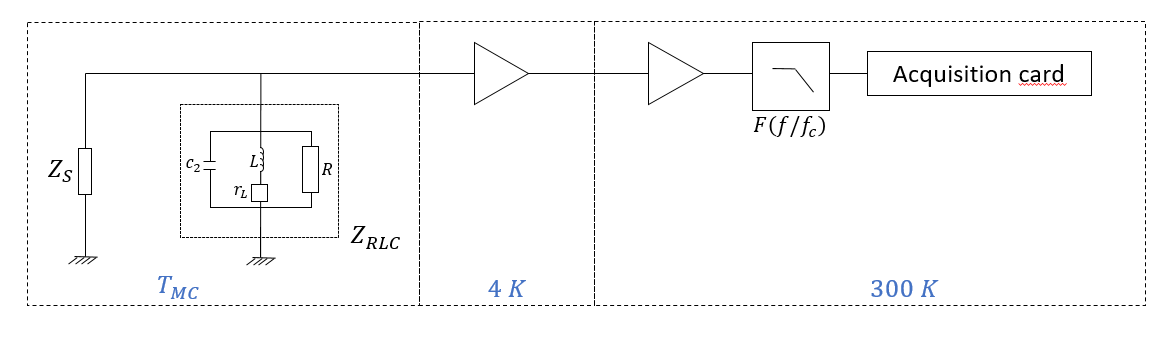}
    \caption{Noise measurement circuit from the sample to the acquisition card. The resonator is made with a capacitor $c_2$, an inductor $L$ with an effective resistance $r_L$ and a resistance $R$ which represents the losses on the circuit. }
    \label{fig:calib_circuit}
\end{figure}

\subsection{Calibration}

\subsubsection{Temperature calibration}

The noise which is measured is the following one :

\begin{equation}
        S^{meas}_{th,v}=G^2 \times \int_{BW} df F( \frac{f}{f_c} ) \left[ S_{v,amp}^2 +
    \left\lvert Z_{//} \right\rvert^2 \left( S^2_{i,amp} +4k_BTRe\left(\frac{1}{Z_{RLC}}\right)+ S^2_{i,sample}(T_S,\Tc=T_S) \right) \right]
    \label{eq:thnoise_meas}
\end{equation}

\begin{figure}[ht!]
    \centering
    \includegraphics[scale=0.70]{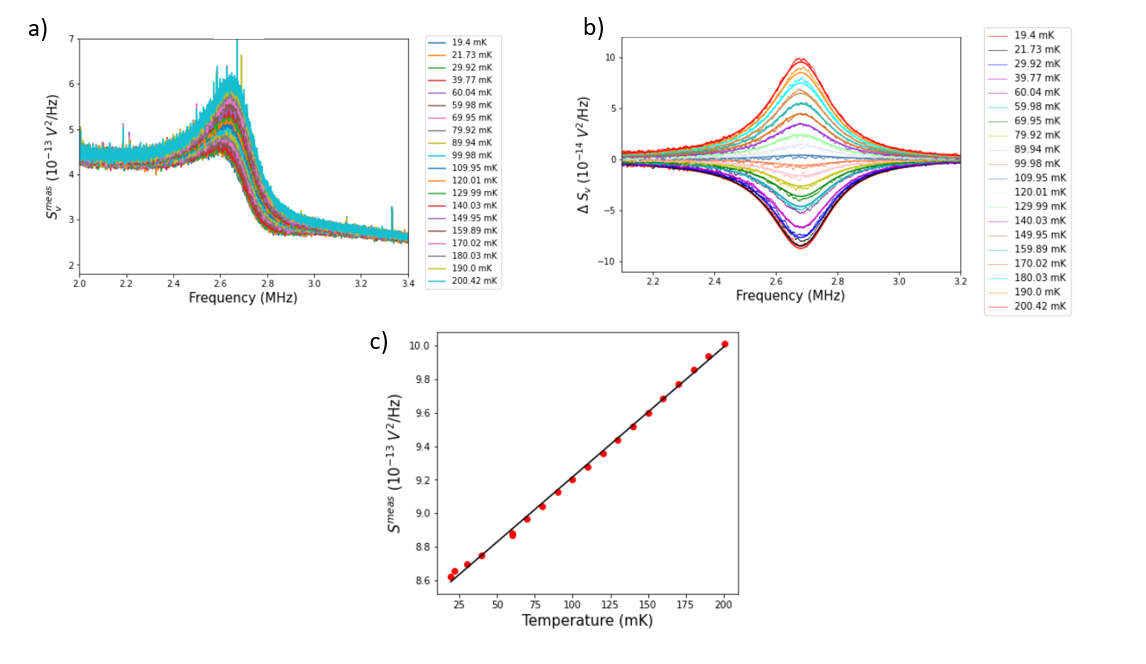}
    \caption{Temperature calibration at 14 T. a)b) Spectrum of the noise with the frequency for different temperature, where a) is direct noise measurement and b)is $\Delta S_v$ with the fitting function (straight lines). The figure c) is the $S^{meas}$ with the temperature of the fridge, where the linear fit is the black line.}
    \label{fig:calib_figures}
\end{figure}

Where $S^2_{i,amp}$ and $S^2_{v,amp}$, are the current and voltage noise of the amplifier, $4k_B T Re(\frac{1}{Z_{RLC}})$ the thermal noise of the resonator which is an LCR circuit, and $S^2_{i,sample}(T_S,\Tc)=3\kB T_S\nu e^2/h+\kB \Tc\nu e^2/h$ the current noise of the sample, where the metallic island can generally be at higher electron temperature $\Tc$ than the rest of the sample, at temperature $T_S$. The unknown are the amplifier noise, and the resonator noise. We determine these parameter with a temperature calibration where we measured the equilibrium noise for temperature ranging between 10 and 200 mK, for various filling factors. Fig.~\ref{fig:calib_figures}a) shows typical raw spectra obtained from this calibration. We first remove the contribution of the temperature-independent terms by calculating the difference between each spectrum and the average of all spectra:
\begin{equation}
    \Delta S_v = S^{meas}_{th}-\left< S^{meas}_{th} \right>_T,
\end{equation}

yielding the curves shown in Fig.~\ref{fig:calib_figures}b), given by the equation, which assumes that all temperatures $T$, $T_S$ and $\Tc$ are equal:

\begin{equation}
    \Delta S^{meas}_v=G^2\int_{BW} F \left(\frac{f}{f_c}\right) 2k_B\Delta T \left\lvert Z_{//} \right\rvert^2 \left[ 2Re \left( \frac{1}{Z_{RLC}} \right) +\nu G_{el}  \right]
    \label{eq:calib_thermalnoise}
\end{equation}

The parameters of the LCR circuit will be found by fitting the above equation \ref{eq:calib_thermalnoise} from the measured noise for a fixed value of $\nu$. The voltage noise of the amplifier can be found from the intercept in temperature dependence of the integrated noise. If we look at the equation \ref{eq:thnoise_meas}, the noise of the amplifier doesn't depend on the temperature. Then, we can do a linear fit, and find  $S_{v,amp_A}\simeq 0.26 \text{ nV /} \sqrt{Hz}$ and $S_{v,amp_B}\simeq 0.28 \text{ nV /} \sqrt{Hz}$.
This calibration can be applied to the measurement, to analyse the current noise of the sample. The calibrated noise is the following :

\begin{multline}\\
    \Delta S = \frac{BW \Delta S^{meas}_v}{G^2 \int_{BW}df F(\frac{f}{f_c}) \left\lvert Z_{//}\right\rvert^2 } \\
    \\
    \Delta S = 2 k_B \nu G_{el} \Delta T \\
\end{multline}

%\subsubsection{Amplifier noise}

%\begin{figure}[ht!]
    %\centering
    %\includegraphics[scale=0.7]{figure_calibration/noiseplateau_allcurves.PNG}
   % \caption{Measure at 3.5T of the average noise for a plateau in a) and for a filling factor in b)}
   % \label{fig:noise_plateau_allcurve}
%\end{figure}

%From the temperature calibration we can only found the value of $S^2_{v,amp}$. The amplifier current noise can be obtained by measured the noise with the temperature, and with the filling factor as show on the figure \ref{fig:noise_plateau_allcurve}. We can rewrite $S^{meas}_v$ from equation \ref{eq:thnoise_meas} with these dependances :

%\begin{equation*}
 % S^{meas}_{th,v}(G_{S}) \propto  \frac{A_{\nu}}{G_{S}^2}+\frac{B_{\nu} }{G_{S}}+C_{\nu} \text{ }  \text{ and }  \text{ } S^{meas}_{th,v}(T) \propto A_T T+B_T
%\end{equation*}

%From these two equations we have :

%\begin{equation*}
   % C_{\nu}=G^2 S_{v,amp}^2
%\end{equation*}

%which gaves $S_{v,amp}$, and

%\begin{equation*}
   %  B_T=G^2 \int_{BW} df F\left(\frac{f}{f_c}\right) \left[S^2_{v,amp}+\left\lvert Z_{//} \right\rvert^2 S^2_{i,amp} \right]
%\end{equation*}

%which gaves $S_{i,amp}$.

%The value of the current and voltage noise found are :

%\begin{table}[h!]
    %\centering
   % \begin{tabular}{|c|c|c|}
    %\hline
   %   & $S_{v,amp_A}$ ( nV$\sqrt{Hz}$ )  & $S_{i,amp}$ ( fA$/\sqrt{Hz}$ )\\
    %  \hline
    %  Line A & 0.26 & 20.7 \\
    %  \hline
   %   Line B & 0.29 & 17.4 \\
   %  \hline
  %  \end{tabular}
  %  \caption{Amplifier noise parameters for both measurement lines.}
  %  \label{tab:amp_param}
%\end{table}

\subsection{Auto and cross-correlations}

We explain here the principle behind our using of auto and cross correlations to extract the temperature increase $\DeltaTc$ from spurious additional contributions. We first begin by establishing the formula linking $\DeltaTc$ to the current noise spectral density $\Delta S$ flowing out of the sample. Following~\cite{Jezouin2013a}, a floating reservoir $\alpha$ at temperature $T_\alpha$ and voltage $V_\alpha$ emits in the $i$-th edge channel flowing out of it a current fluctuation given by:

\begin{equation}
    \delta I_i=\delta I_i^{T_\alpha}+G_0\delta V_\alpha
    \label{eq:currfluccontact}
\end{equation}

The first term $\delta I_i^{T_\alpha}$ corresponds to current fluctuations due to the finite temperature of the reservoir, with a spectral density $<(\delta I_i^{T_\alpha})^2>=2G_0 k_B T_\alpha$; the second term corresponds to fluctuations of the voltage of the floating reservoir. Importantly, in each of the edge channels, the first term is uncorrelated, $<\delta I_i^{T_\alpha}\delta I_j^{T_\alpha}>=\delta_{i,j}\times2G_0 k_B T_\alpha$, while the second is correlated, \textit{i.e.} it is equal at all times in all edge channels flowing out of the reservoir.
We adopt the following notations for the contacts: $\alpha=c$ for the central floating metallic island, $\alpha=A,B$ for the contacts connected to noise measurement lines with complex input impedances $Z_{A,B}$, and $\alpha=Ain,Bin$ for the current feed contacts upstream of the floating island on the $A,B$ side, the voltages of which are assumed to be without fluctuations. Assuming that we are in the integer QH regime with filling factor $\nu$, the current balances using Eq.~\ref{eq:currfluccontact} on contacts $A$, $B$ and $c$ read:

\begin{eqnarray}
\delta V_A (\nu G_0+1/Z_A)+\sum_i \delta I_i^{T_A}=\left(\sum_i \delta I_i^{T_c}\right)^A+\nu G_0\delta V_c\\
\delta V_B (\nu G_0+1/Z_B)+\sum_i \delta I_i^{T_B}=\left(\sum_i \delta I_i^{T_c}\right)^B+\nu G_0\delta V_c\\
\left(\sum_i \delta I_i^{T_c}\right)^A+\left(\sum_i \delta I_i^{T_c}\right)^B+2\nu G_0\delta V_c=\sum_i \delta I_i^{T_{Ain}}+\sum_i \delta I_i^{T_{Bin}}
    \label{eq:currflucbal}
\end{eqnarray}

The thermal noise of the measurement impedances $Z_{A,B}$ is neglected here for simplicity, and $\left(\sum_i \delta I_i^{T_c}\right)^{A/B}$ corresponds to the sum of the thermal current fluctuations flowing from the metallic island to contacts $A/B$. We combine these equations to express the voltage fluctuations $\delta V_A$ and $\delta V_A$ as a function of all other current fluctuations:

\begin{eqnarray}
\delta V_A (\nu G_0+1/Z_A)=-\sum_i \delta I_i^{T_A}+\frac{1}{2}\left(\sum_i \delta I_i^{T_c}\right)^A-\frac{1}{2}\left(\sum_i \delta I_i^{T_c}\right)^B+\frac{1}{2}\sum_i \delta I_i^{T_{Ain}}+\frac{1}{2}\sum_i \delta I_i^{T_{Bin}}\\
\delta V_B (\nu G_0+1/Z_B)=-\sum_i \delta I_i^{T_B}-\frac{1}{2}\left(\sum_i \delta I_i^{T_c}\right)^A+\frac{1}{2}\left(\sum_i \delta I_i^{T_c}\right)^B+\frac{1}{2}\sum_i \delta I_i^{T_{Bin}}+\frac{1}{2}\sum_i \delta I_i^{T_{Ain}}
    \label{eq:volflucAB}
\end{eqnarray}

One can thus see right away that the terms containing thermal fluctuations of the metallic islands are anti-correlated, while the thermal fluctuations of $A/B$ are uncorrelated and those of $Ain/Bin$ are positively correlated. Assuming first that $T_A=T_B=T_{Ain}=T_{Bin}=T_0$ (that is, only the metallic island heats up while all other contacts stay at base electron temperature), we can calculate the auto and crosscorrelated voltage noise spectra:

\begin{eqnarray}
    <(\delta V_A)^2>=\frac{1}{|\nu G_0+1/Z_A|^2}\left[ 3\nu G_0 \kB T_0 + \nu G_0 \kB T_c \right]\\
    <(\delta V_B)^2>=\frac{1}{|\nu G_0+1/Z_B|^2}\left[ 3\nu G_0 \kB T_0 +\nu G_0 \kB T_c\right]\\
    <\delta V_A(\delta V_B)^\ast> =\frac{1}{(\nu G_0+1/Z_A)(\nu G_0+1/Z_B)^\ast}\left[ \nu G_0 \kB T_0 - \nu G_0 \kB T_c \right]
    \label{eq:volcorrAB}
\end{eqnarray}

This can finally be expressed as a function of the excess thermal current spectrum $\Delta S=\nu G_0 \kB \Tc$:

\begin{eqnarray}
    <(\delta V_A)^2> =\frac{1}{|\nu G_0+1/Z_A|^2}\left[ 4\nu G_0 \kB T_0 + \Delta S\right]\\
    <(\delta V_B)^2>=\frac{1}{|\nu G_0+1/Z_B|^2}\left[ 4\nu G_0 \kB T_0 + \Delta S\right]\\
    <\delta V_A(\delta V_B)^\ast> =-\frac{1}{(\nu G_0+1/Z_A)(\nu G_0+1/Z_B)^\ast}\times\Delta S
    \label{eq:volcorrABdeltaS}
\end{eqnarray}

\begin{figure}[ht]
\centering
\includegraphics[width=0.8\textwidth]{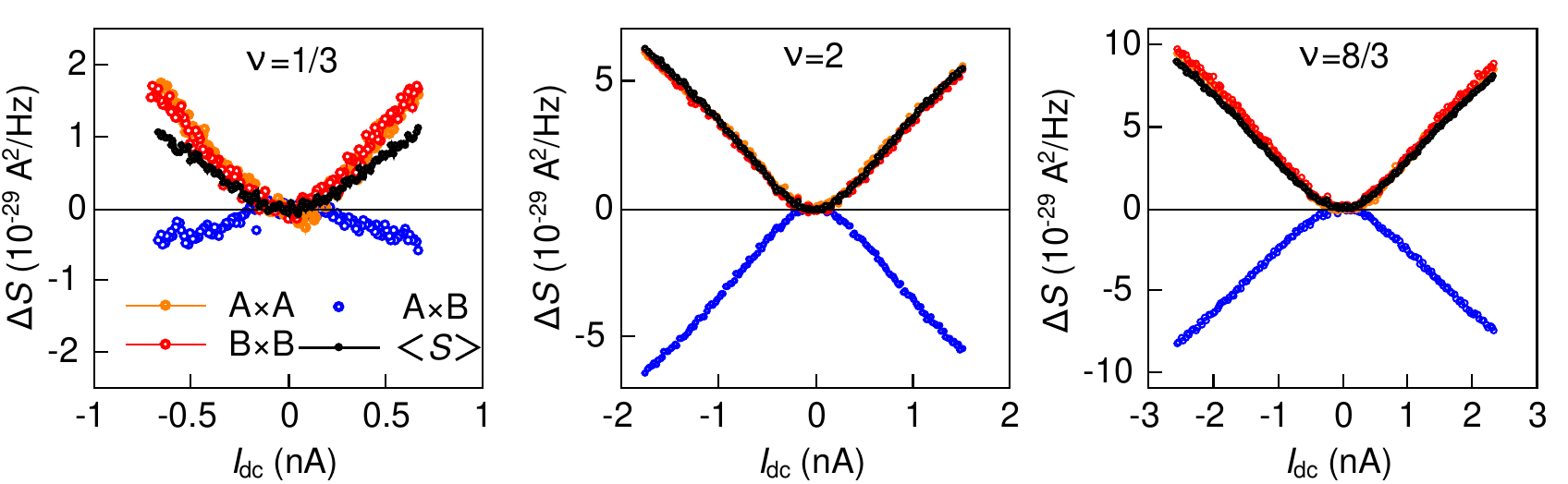}
\caption{\label{Figsm-autovscrossCD1} Auto- (orange and red symbols) and crosscorrelation (blue symbols), as well as the computed thermal contribution $<S>$ (black symbols), as a function of $\Idc$, for cooldown 1 at filling factors $\nu=1/3$ (left), $2$ (middle), and $8/3$ (right).}
\end{figure}

\begin{figure}[ht]
\centering
\includegraphics[width=0.35\textwidth]{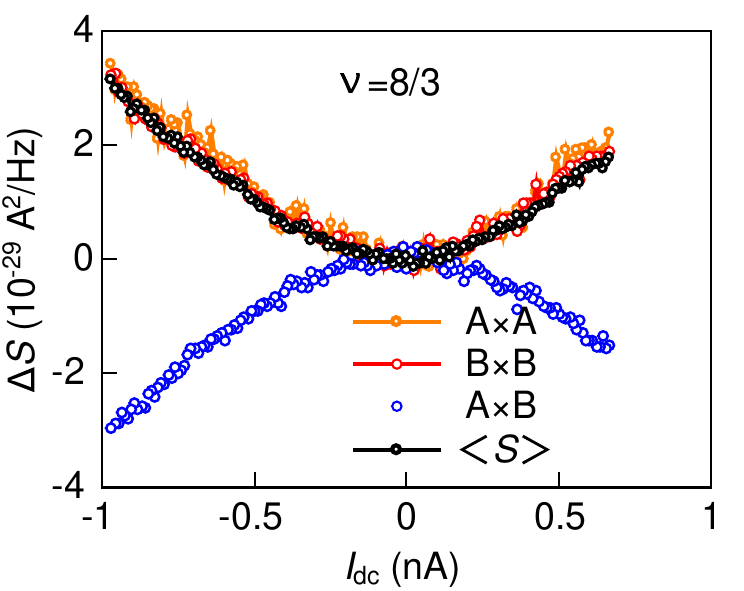}
\caption{\label{Figsm-autovscrossCD2} Auto- (orange and red symbols) and crosscorrelation (blue symbols), as well as the computed thermal contribution $<S>$ (black symbols), as a function of $\Idc$, for cooldown 2 at filling factor $\nu=8/3$.}
\end{figure}

Thus, an increase of the metallic island electron temperature $\DeltaTc$ leads to positive autocorrelations, and negative crosscorrelation. Following \cite{Sivre2019}, we rely on this to separate the contribution of thermal noise due to the increase of the metallic island's electron temperature from additional spurious sources of noise. These sources can have three origins: current fluctuations generated upstream of the island, after the island, and at the island interface. In the previous calculations, the first two cased can be encompassed by introducing effective temperatures: $T_{Ain/Bin}$ for the upstream noise, and $T_{A/B}$ for the noise downstream of the island. Recalling the above equations, this leads to:

\begin{eqnarray}
    <(\delta V_A)^2>=\frac{1}{|\nu G_0+1/Z_A|^2}\left[ 2\nu G_0 \kB T_A + \nu G_0 \kB T_c +\frac{1}{2}\nu G_0 \kB T_{Ain} + \frac{1}{2}\nu G_0 \kB T_{Bin}\right]\\
    <(\delta V_B)^2>=\frac{1}{|\nu G_0+1/Z_B|^2}\left[ 2\nu G_0 \kB T_B + \nu G_0 \kB T_c +\frac{1}{2}\nu G_0 \kB T_{Bin} + \frac{1}{2}\nu G_0 \kB T_{Ain}\right]\\
    <\delta V_A(\delta V_B)^\ast> =\frac{1}{(\nu G_0+1/Z_A)(\nu G_0+1/Z_B)^\ast}\left[- \nu G_0 \kB T_c +\frac{1}{2}\nu G_0 \kB T_{Bin} + \frac{1}{2}\nu G_0 \kB T_{Ain}\right]
    \label{eq:volcorrABdifftemp}
\end{eqnarray}

As described in the main text, an additional upstream noise increases both auto and crosscorrelations by the same amount, which can be subtracted by computing the difference between auto and crosscorrelation. Conversely, noise downstream of the island will lead to different autocorrelations. In that case, it becomes problematic to extract the thermal contribution, as the added noise cannot be easily subtracted. We thus discard the data as no reliable heat transport analysis can be performed.

The case of current fluctuations generated at the island interface requires additional analysis. We present here a simple model based on a single edge channel for simplicity. The imperfect interface is modelled by a scatterer with transmission $\tau$ inserted before the island. We assume here that only one interface (the one connected to the A side) is imperfect, such that the scatterer reflects the current stemming from contact $Ain$ to contact $A$ with a coefficient $1-\tau$. The scatterer generates additional shot noise noted $\delta I^\ast$. Current conservation implies that the fluctuation thusly generated on either side of the scatterer is opposite: by convention, we write $-\delta I^\ast$ the current fluctuation emitted in the edge channel flowing to $A$, and $+\delta I^\ast$ the fluctuation emitted in the edge channel going to the island. This changes the current balances shown in Eq.~\ref{eq:currflucbal} to the following:

\begin{eqnarray}
\delta V_A (G_0+1/Z_A)+\delta I^{T_A}=-\delta I^\ast+(1-\tau)\delta I^{T_Ain}+\tau\left( \delta I^{T_c}\right)^A+\tau G_0\delta V_c \\
\delta V_B (G_0+1/Z_B)+ \delta I^{T_B}=\left( \delta I^{T_c}\right)^B+G_0\delta V_c\\
(1+\tau)G_0\delta V_c+\tau\left(\delta I^{T_c}\right)^A+\left(\delta I^{T_c}\right)^B=\delta I^\ast+\tau\delta I^{T_{Ain}}+\delta I^{T_{Bin}}
    \label{eq:currflucbalshotnoise}
\end{eqnarray}

Expressing, as above, $\delta V_{A/B}$ as a function of all current fluctuations yields (noting $F(\tau)=\tau/(1+\tau)$):

\begin{eqnarray}
\delta V_A (G_0+1/Z_A)=-\delta I^{T_A}-\frac{F(\tau)}{\tau}\delta I^\ast+F(\tau)\left(\delta I^{T_c}\right)^A-F(\tau)\left( \delta I^{T_c}\right)^B+\left[ 1-\tau+F(\tau)\right]\delta I^{T_{Ain}}+F(\tau) \delta I^{T_{Bin}}\\
\delta V_B (G_0+1/Z_B)=-\delta I^{T_B}+\frac{F(\tau)}{\tau}\delta I^\ast-F(\tau)\left(\delta I^{T_c}\right)^A+\frac{F(\tau)}{\tau}\left( \delta I^{T_c}\right)^B+F(\tau)\delta I^{T_{Ain}}+\frac{F(\tau)}{\tau} \delta I^{T_{Bin}}\\
    \label{eq:volflucABshotnoise}
\end{eqnarray}

Thus, the contribution of the shot noise $<(\delta I^\ast)^2>$ is the same for auto and crosscorrelations, with a negative sign for the cross correlations. While this is quite similar to the contribution of the thermal noise for a perfect interface, it turns out that this same thermal noise has now different contributions in the two autocorrelations. This can be intuitively understood by the fact that the side with the imperfect interface sees less noise stemming from the island. In the end, by using the equal autocorrelation criterion mentioned above, one can again make sure that the noise measured only stems from the increase island temperature $T_c$.

 To sum up, we always discard datasets with differing autocorrelations, as it signals shot noise at either the island interface, or downstream of it, and cannot be easily subtracted. We then extract the thermal noise contribution by computing the averaged difference between autocorrelation and crosscorrelation $<S>=((A\times A+B \times B)/2-A\times B)2$. Systematically calculating $<S>$ (provided, again, that the autocorrelations are equal) has the added benefit of increasing our signal to noise ratio. In the main text, all $\Delta \Tc$ are extracted from the computed $<S>$. We show examples of the procedure in Fig.~\ref{Figsm-autovscrossCD1} for cooldown 1 at $\nu=1/3$, $2$, and $8/3$, as well as in Fig.~\ref{Figsm-autovscrossCD2} for cooldown 2 at $\nu=8/3$. The autocorrelations are shown as red and orange symbols, the crosscorrelation as blue symbols, and $<S>$ as black symbols. We generally observe that there is no additional noise for integer $\nu$, while fractional $\nu$ tend to display the presence of additional noise. In particular, the additional noise at $\nu=1/3$ for cooldown 1 is substantial, being comparable in amplitude to the thermal noise contribution. Note that other fractional filling factors, such as $\nu=2/3$, or $\nu=1/3$ for cooldown 2, showed significantly different autocorrelation and were not considered in the heat transport analysis. Similarly, at large bias, the autocorrelation at $\nu=8/3$ for cooldown 2 started differing notably; for this reason we restricted ourselves to biases smaller than $1~$nA for this dataset (see also Fig.~\ref{figsm_TvsInu83} below).

Note that a poor interface at the floating island will also impact the conductance measurements, with the reflected differential transconductance being larger than the transmitted one. Checking that the transconductance are equal and quantized is thus crucial in these measurements.
Interestingly, a bad transparency of the injection contact upstream of the floating island will not necessarily appear in the conductance measurement, as the sample is current biased (an imperfect transparency will reduce the voltage drop on the injection contact, and add noise, but won’t change the amount of current fed flowing from this contact). Finally, our sample design is such that the floating island has a larger interface length than the other contacts (which is usually the opposite in GaAs thermal transport experiments). Thus, poor interfaces are more likely to stem from these contacts.

Note that the presence of additional noise, and its amplitude, is not universal. For instance, in ref.~\cite{Srivastav2022}, the data at $\nu=1/3$ does not seem to present spurious noise, as the authors only relied on autocorrelations to extract $\Tc$. This is likely due to edge contacts, which are known to locally dope graphene in their vicinity~\cite{Wei2018}, and the microscopics of which largely depends on the recipes used during the samples' fabrication.

\subsection{$\nu=8/3$ data}

\begin{figure}[ht]
\centering
\includegraphics[width=0.35\textwidth]{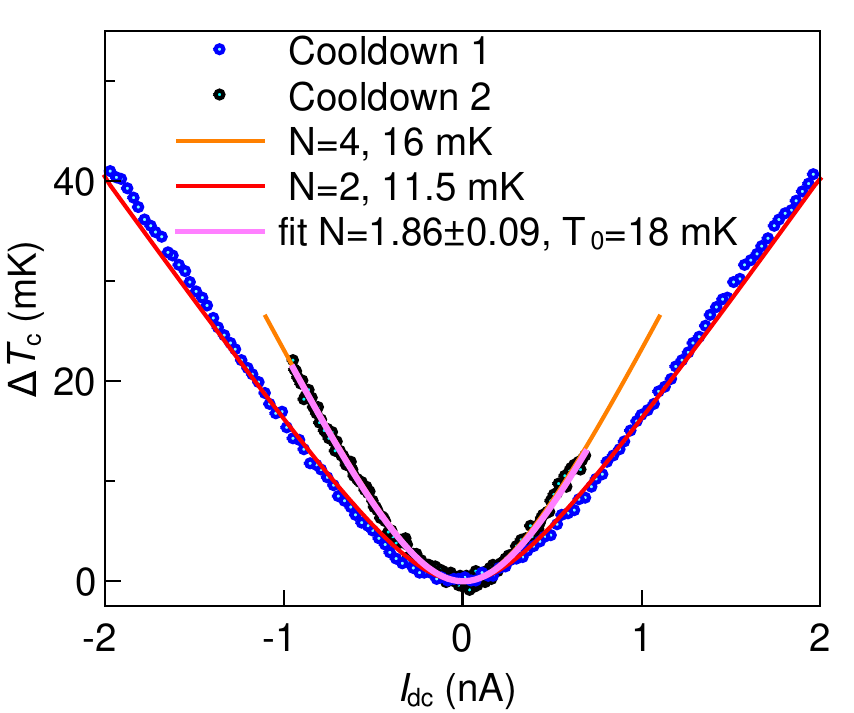}
\caption{\label{figsm_TvsInu83} $\DeltaTc$ versus $\Idc$ at $\nu=8/3$, for cooldowns 1 (blue circles) and 2 (black) circles. The lines are heat transport model fits: red/orange: fits with $N$ fixed; pink: fit of cooldown 2 with both $T_0$ and $N$ as free parameters.}
\end{figure}

Fig.~\ref{figsm_TvsInu83} shows the $\DeltaTc(\Idc)$ data at $\nu=8/3$ for both cooldowns discussed in the main text. As mentioned in the text, both measurements are strikingly different, with the second cooldown data showing a significantly higher slope, corresponding to a smaller $N$. The lines are heat transport model fits, detailed below.

\subsection{Base electron temperature $T_0$}
\subsubsection{Extraction from heat balance fits}

\begin{figure}[ht]
\centering
\includegraphics[width=0.65\textwidth]{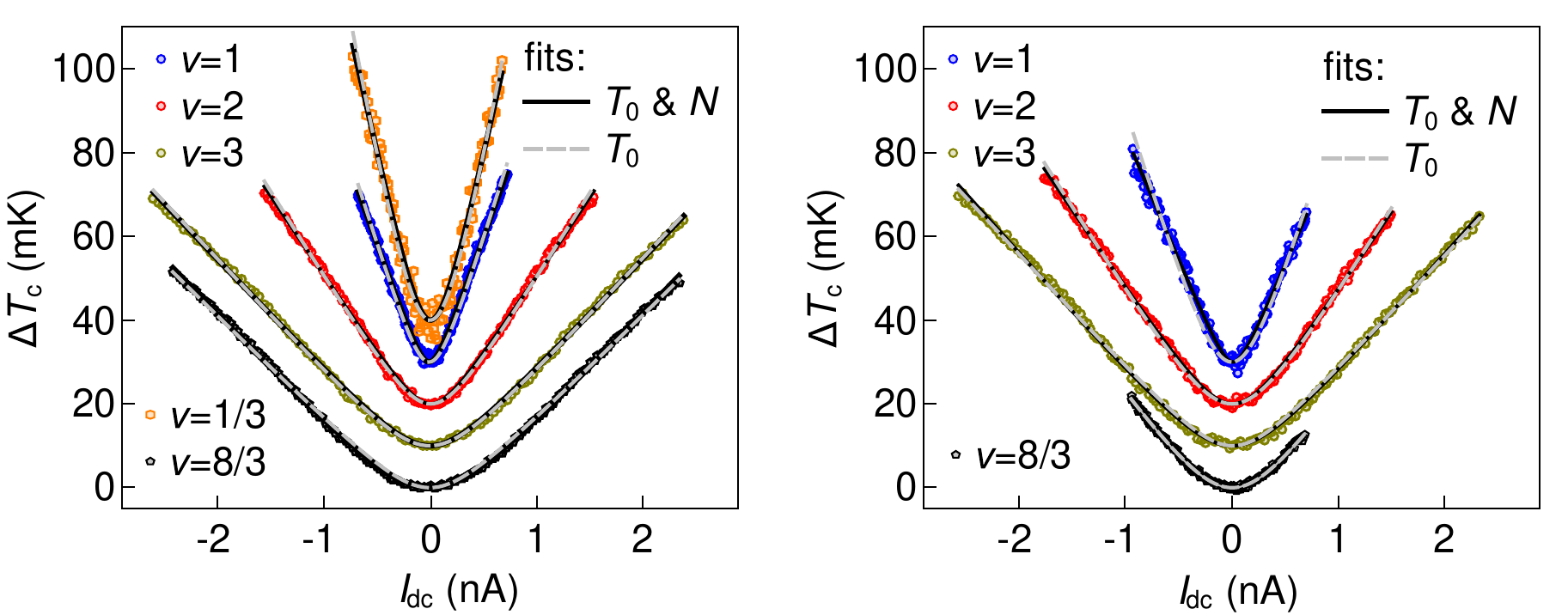}
\caption{\label{figsm-TvsI_fits} $\DeltaTc$ versus $\Idc$, measured for cooldown 1 (left) and cooldown 2 (right). The symbols are the experimental data shown in the main text. The grey dashed line are fits using the heat transport model with fixed $N$, also shown in the main text. The black lines are fit with both $N$ and $T_0$ as free parameters.}
\end{figure}

Fig.~\ref{figsm-TvsI_fits} shows fits of $\DeltaTc$ versus $\Idc$ data discussed in the main articles with the heat transport model, with either $N$ fixed and $T_0$ a free parameter (grey dashed lines, see also main text), or with both $N$ and $T_0$ as free parameters (black lines). The agreement between either fits and the data is generally good; comparing the extrated values between the two fits allows us to infer the limits and the precision of our approach. These values are summarized in Table~\ref{tab:fits_param}.

\begin{table}[h!]
    \centering
    \begin{tabular}{|c|c|c|c|c|c|}
    \hline
      &  $\nu=1$  & $\nu=2$ & $\nu=3$ &  $\nu=1/3$  & $\nu=8/3$ \\
      \hline
      CD 1 - $N$ fixed &  $T_0=11~$mK  & $T_0=11~$mK & $T_0=12~$mK &  $T_0=42~$mK  & $T_0=12~$mK \\
      \hline
      CD 1 - $N$ free  &  $N=1.4$, $T_0=7.4~$mK  & $N=2.3$, $T_0=8.7~$mK & $N=3.1$, $T_0=10~$mK &  $N=1.5$, $T_0=24~$mK  & $N=3.5$, $T_0=14~$mK \\
       \hline
      CD 2 - $N$ fixed  &  $T_0=20~$mK  & $T_0=15~$mK & $T_0=13~$mK &    & $T_0=16~$mK \\
       \hline
      CD 2 - $N$ free  &  $N=1.9$, $T_0=7.0~$mK  & $N=2.4$, $T_0=13~$mK & $N=2.7$, $T_0=13~$mK &    & $N=1.9$, $T_0=18~$mK \\
      \hline
    \end{tabular}
    \caption{Heat transport parameters extracted from the fits shown in Fig.~\ref{figsm-TvsI_fits}. CD means cooldown.}
    \label{tab:fits_param}
\end{table}

The first observation is that leaving $N$ as a free parameter generally leads to an overestimation of $N$, that is particularly pronounced at $nu=1/3$ and $\nu=1$. The associated $T_0$ are generally smaller than for fixed $N$. In particular, the $T_0\approx7~$mK obtained at $\nu=1$ are not physical, being lower than the calibrated base temperature of our refrigerator of about $8.7~$mK. The systematic overestimation of $N$ likely stems from the fact that the fitting procedure puts an important weight on the largest bias data, where the temperature increase is such that electron-phonon coupling starts playing a role and in effect increases the outgoing heat flow. In order to obtain a good fit of not only the slope, but the absolute value of the data, the procedure in turn underestimates $T_0$, which increases the absolute value of the fit at finite bias. Except for the most problematic values at low filling factors, comparing both fitting procedures allows fixing an upper bound for our determination of $N$ of about $0.2-0.3$. 

\subsubsection{Equilibrium noise}

\begin{figure}[ht]
\centering
\includegraphics[width=0.55\textwidth]{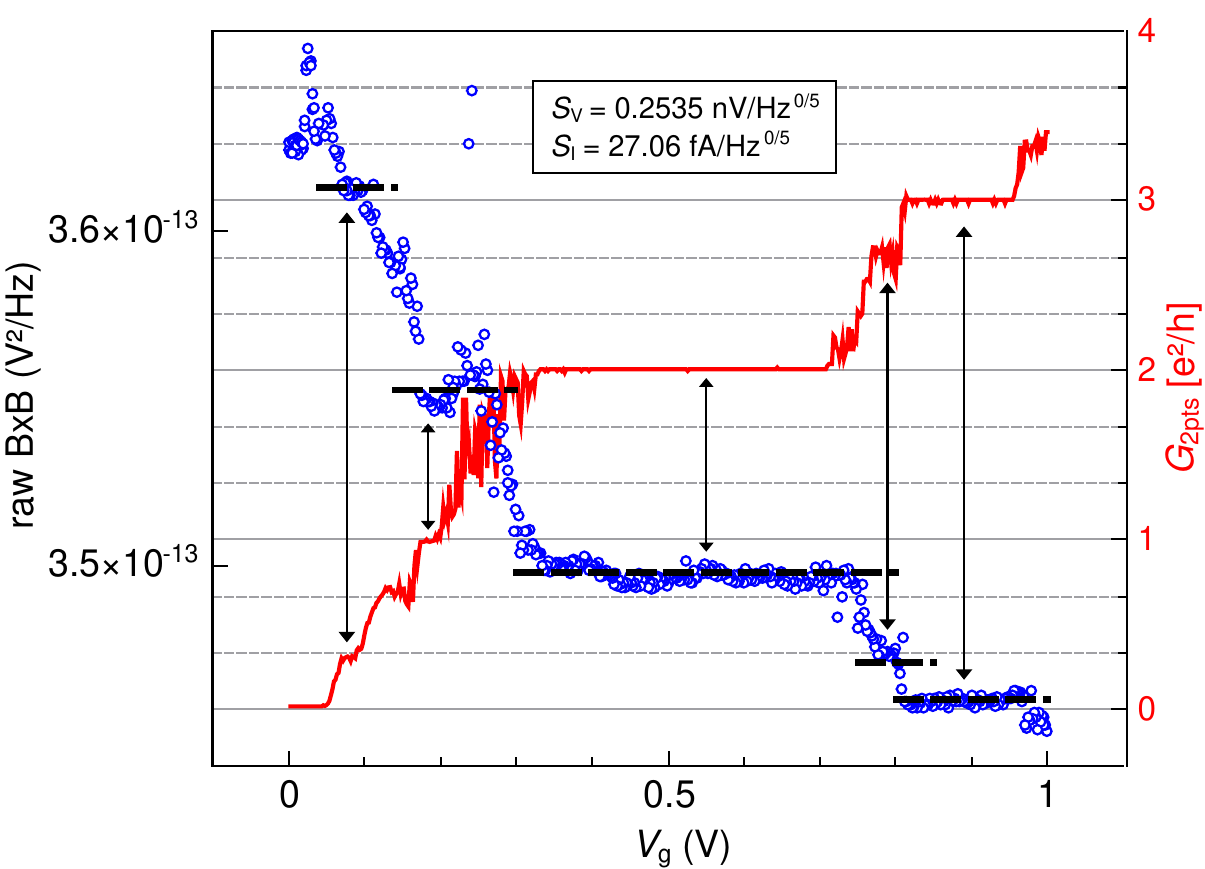}
\caption{\label{figsm-noiseplateauCD1} Raw noise on line B (blue circles) and $G_{2\mathrm{pt}}$ (red) versus $\Vg$ for cooldown 1. The position of the quantum Hall plateaus are indicated by the vertical arrows. The horizontal dashed lines correspond to the computed value of the noise using the calibration data and the extracted $T_0$.}
\end{figure}

\begin{figure}[ht]
\centering
\includegraphics[width=0.8\textwidth]{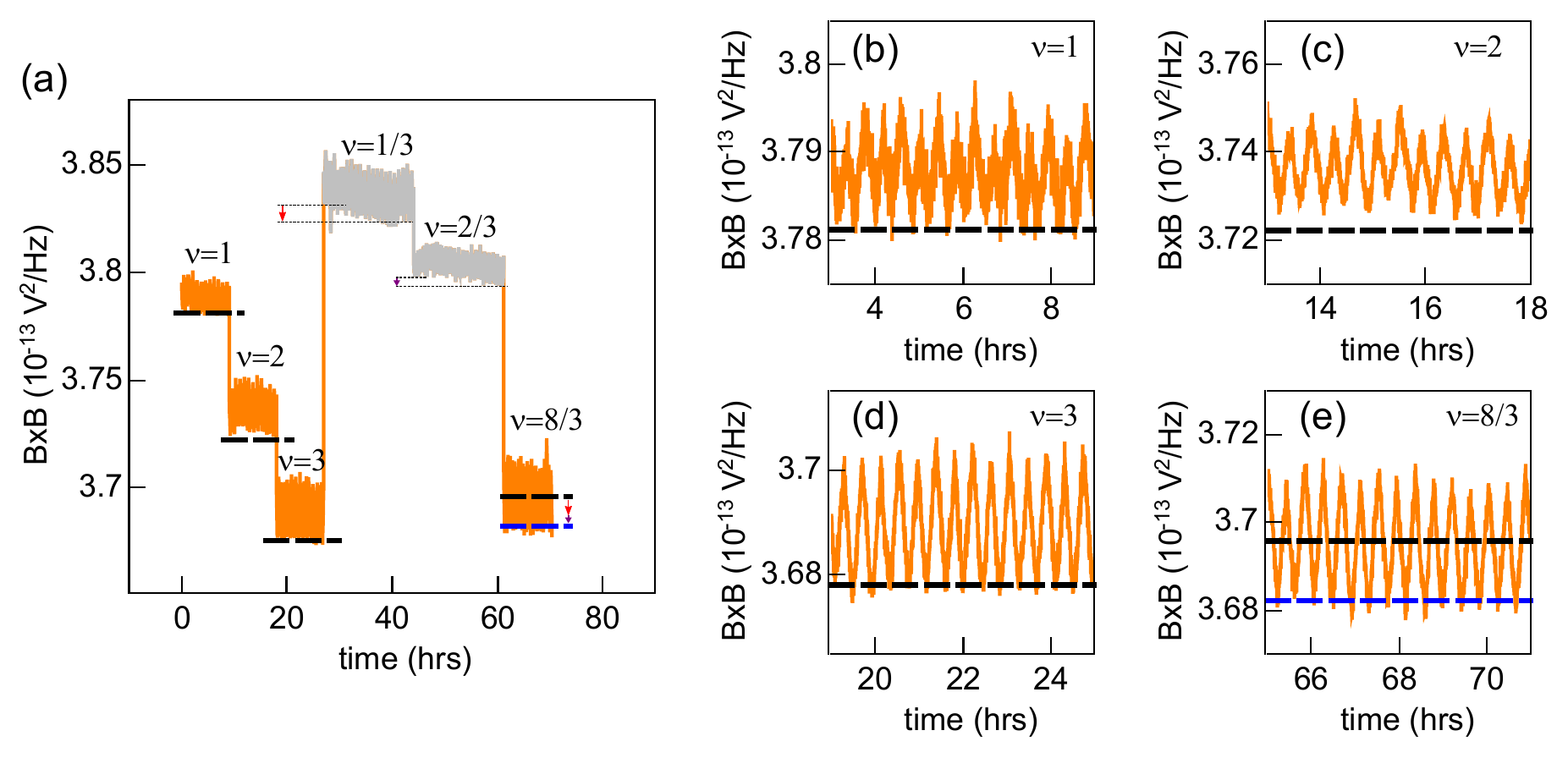}
\caption{\label{figsm-noiseplateauCD2} \textbf{a),} raw noise on line B versus time, during the measurement run in cooldown 2 corresponding to the heat transport data shown in the main text. The greyed out data, corresponding to $\nu=1/3$ and $2/3$, is not discussed in the main text as it did not satisfy the criterion of equal autocorrelations. The black horizontal dashed lines correspond to the computed value of the noise using the extracted $T_0$. The blue horizontal dashed line corresponds to the computed value of the noise, after taking into account the slow noise drift occurring during the $\nu=1/3$ and $2/3$ measurements (blue and red vertical arrows).}
\end{figure}
The values of $T_0$ extracted from the heat balance fits can be compared with the amplitude of the noise measured at equilibrium (that is, $\Idc=0$). Indeed, at equilibrium the raw value of the noise $S^{meas}_{th,v}$ for a given line is given by supplementary Eq.~\ref{eq:thnoise_meas}. We thus compute $S^{meas}_{th,v}$ with the $T_0(\nu)$ extracted from the fits, and compare it to the equilibrium noise. The latter can be obtain by different ways. First, one can directly measure the noise at $\Idc=0$ as a function of $\Vg$, as illustrated in Fig.~\ref{figsm-noiseplateauCD1} for the first cooldown, showing plateaus corresponding to the different filling factors, the value of which can be reproduced with supplementary Eq.~\ref{eq:thnoise_meas}. Second, one can look at the raw noise data obtained for the heat transport measurement, and match the value obtained at zero bias with supplementary Eq.~\ref{eq:thnoise_meas}. This is illustrated in Fig.~\ref{figsm-noiseplateauCD2} for the second cooldown; note that this second method is much more sensitive to the slow drifts in the noise signals, that typically occur over several hours of measurement. In particular, Fig.~\ref{figsm-noiseplateauCD2} shows how the slow drifts occurring during the measurements at $\nu=1/3$ and $2/3$, indicated by the vertical red and blue arrows, have to be taken into account to match the equilibrium noise for the last measurement at $\nu=8/3$. With both methods, we obtain a very reasonable agreement.

%\begin{equation}
%V\times V=G^2 \int_{f_1}^{f_2} \frac{df}{\Delta f} \left( \frac{1}{1+(f/f_c)^2} \right) ^2\left[ S_v^2 +\left | \frac{1}{1/Z_\mathrm{res}(f)+1/R_\mathrm{s}} \right |^2 \left (S_i^2 + 4 \kB T_\mathrm{res} \Re \left [\frac{1}{Z_\mathrm{res}(f)} \right] + 4 \kB T_\mathrm{s}\frac{1}{R_\mathrm{s}}\right) \right],
%\label{eq:SM-eqnoise}
%\end{equation}

%where the gain $G$, the resonator impedance $Z_\mathrm{res}(f)$ and the cutoff frequency $f_c$ are extracted from the calibration described above, $R_s=h/\nu e^2$ is the 2-point resistance of the sample seen from the noise measurement line, $T_\mathrm{res}$ and $T_\mathrm{s}$ are respectively the resonator and the sample electron temperature, and $S_v$ and $S_i$ are the cryogenic amplifier's voltage and current input noise.

\subsubsection{Linearity of heat flow versus $T^2$}

\begin{figure}[ht]
\centering
\includegraphics[width=0.8\textwidth]{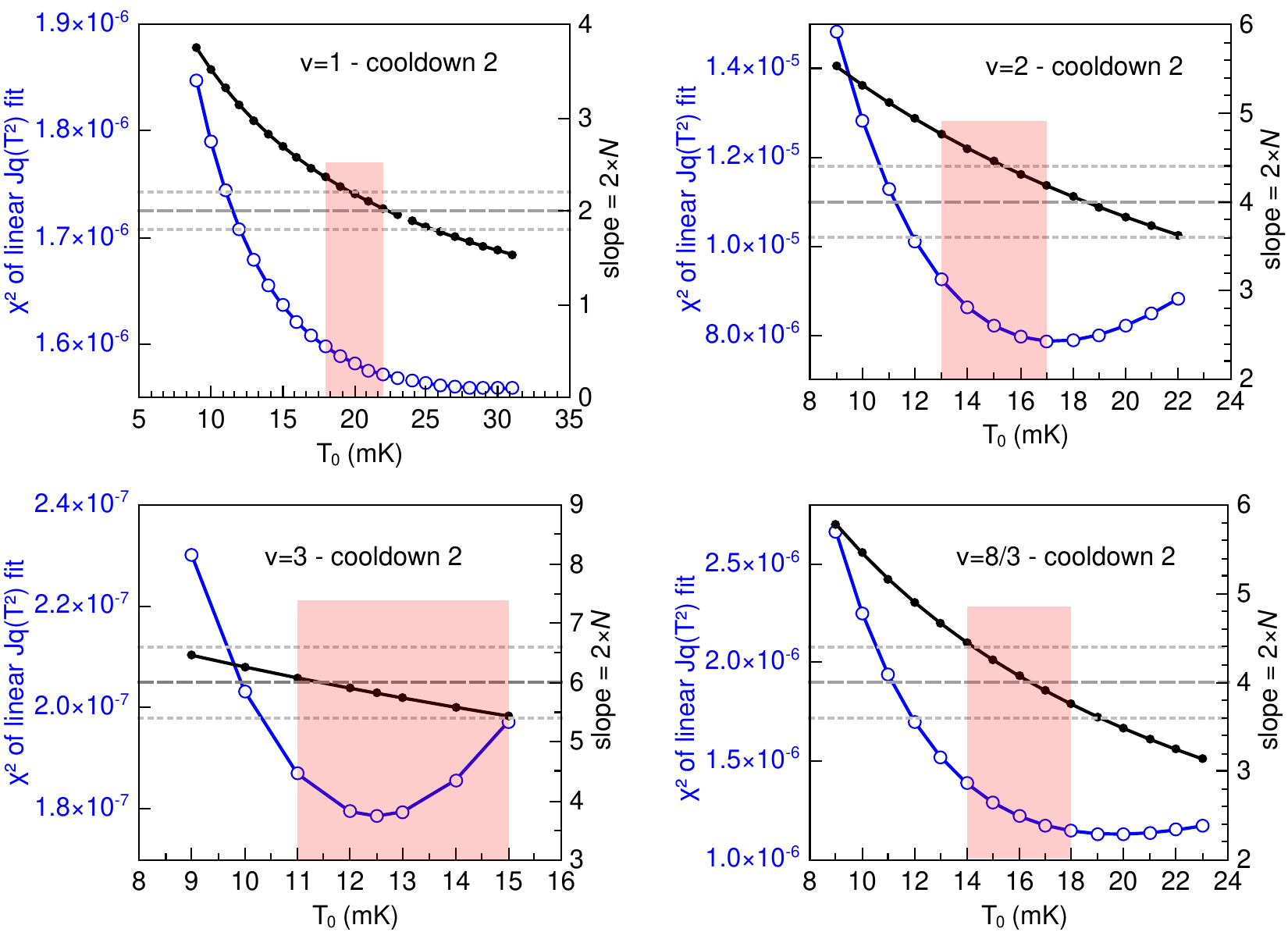}
\caption{\label{figsm-chi2} $\chi^2$ extracted from linear fits of $\Jq(\Tc^2-T_0^2)$ as a function of the assumed value $T_0$, for various filling factors in cooldown 2. Blue symbols and left Y-axis: $\chi^2$. Black dots: corresponding slope $2N$. The horizontal thick dashed line corresponds to the expected value of $2N$, and the thin dotted lines to a $\pm10~\%$ deviation. The red area corresponds to $T_0$ in a $\pm2~$mK range of the $T_0$ extracted from the fits shown in the main text.}
\end{figure}

A further check performed to ensure our confidence in the extraction of $T_0$ consists in quantitatively inferring for which $T_0$ the data, expressed in terms of quantized heat flow (that is, by plotting $\Jq$ versus $\Tc^2-T_0^2$), is the closest to a linear behavior, and to which value of $N$ this corresponds. The advantage of this approach is that it allows us putting a stronger emphasis on the lowest bias data, where the electron-phonon coupling plays no role. We illustrate it in Fig.~\ref{figsm-chi2}, where we have plotted the $\chi^2$ of systematic linear fits of the $\Jq(\Tc^2-T_0^2)$ data for different values of $T_0$, for $\nu=1$, 2, 3 and $8/3$ in cooldown 2. We observed that the $\chi^2$ is generally minimal for values of $T_0$ close to the ones extracted using the fits shown in the main text, with corresponding values of $N$ in good agreement with our expectations.

\subsection{Electron-phonon cooling}

\begin{figure}[ht]
\centering
\includegraphics[width=0.5\textwidth]{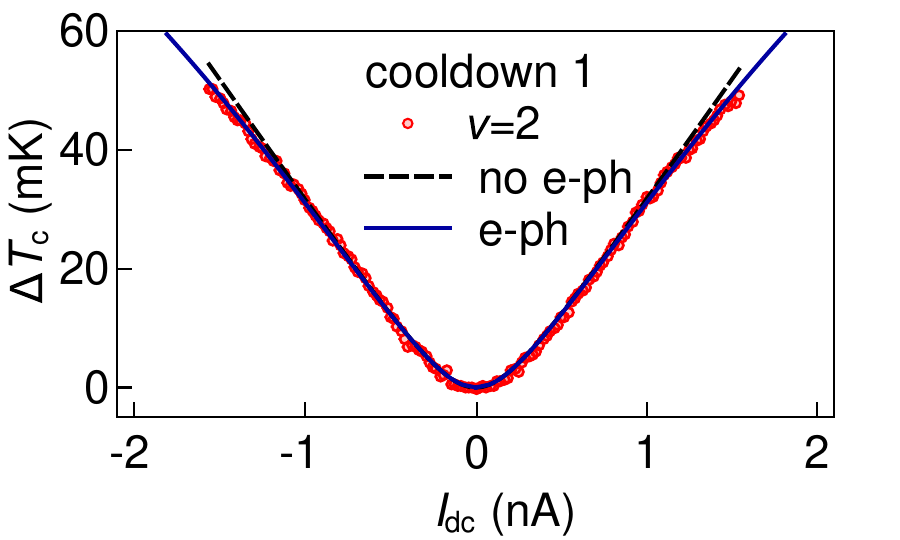}
\caption{\label{figsm-eph-nu2cd1} $\DeltaTc$ versus $\Idc$ at $\nu=2$. Symbols: experimental data for cooldown 1, shown in the main text. Black dashed line: heat transport fit with $N=4$ fixed and $T_0=11~$mK (see main text). Dark blue line: heat transport model including electron-phonon coupling, with $T_0=11~$mK.}
\end{figure}

In addition to heat transport due to the integer and fractional quantum Hall edge modes, the Joule heat dissipated in the metallic island can also be evacuated through the electron-phonon coupling in the island. This cooling contribution is usually written as $J_\mathrm{Q}^\mathrm{e-ph}=\Sigma \Omega\times(\Tc^\delta-T_0^\delta)$, with $\Sigma$ a constant depending on the material, $\Omega$ the volume of the metallic island, and $\delta\approx5$ \cite{Jezouin2013a}. Here, the phonons are assumed to be thermalized at the same base temperature $T_0$ as the equilibrium electrons (note that this might not always be the case, especially for low filling factors where the extracted $T_0$ is higher). The effect of this contribution is to bring a sublinear behavior of $\DeltaTc$ versus $\Idc$ at large $\Idc$, when the temperature increase becomes large enough; this sublinearity becomes all the more pronounced for increased $T_0$ or large volumes of the metallic island. In our case, the low $T_0$ and the reduced size of the metallic island minimize this contribution. Still, small sublinearities can be observed, and accounted for by adding the electron-phonon coupling contribution to the model. This is exemplified in Fig.~\ref{figsm-eph-nu2cd1}, which shows the $\nu=2$ data for cooldown 1, along with heat transport models with and without electron-phonon coupling. The latter case, with the same $T_0\approx11~$mK, $\delta=5$ and $\Sigma\Omega\approx0.8~$nW$/$K$^{-5}$, reproduces well the small sublinearity in the data. This being said, the effect is small enough to be safely dismissed in our analysis.

\subsection{Heat Coulomb blockade}

\begin{figure}[ht]
\centering
\includegraphics[width=0.5\textwidth]{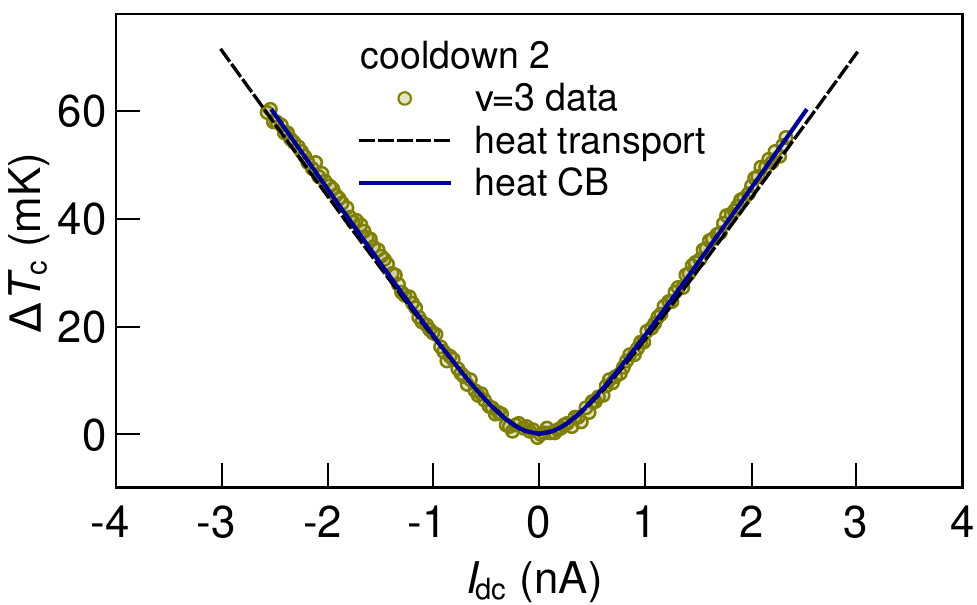}
\caption{\label{figsm_heatCBnu3-CD2} $\DeltaTc$ versus $\Idc$ at $\nu=3$. Symbols: experimental data for cooldown 2, shown in the main text. Black dashed line: heat transport fit with $N=3$ fixed and $T_0=12.5~$mK (see main text). Dark blue line: heat Coulomb blockade model, with $T_0=15~$mK.}
\end{figure}

The range of base electronic temperatures attained in our experiment raises the question whether we are sensitive to heat Coulomb blockade, a recently observed mechanisms wherein heat transport is reduced by one channel due to the inability to change the metallic island's charge state at energy and temperature much lower than its charging energy $E_\mathrm{c}$ \cite{Sivre2018}. This effect is challenging to observed in its fully developed (that is, one ballistic heat transport channel being fully suppressed), typically requiring $\kB T\sim0.02 E_\mathrm{c}$ \cite{Sivre2018}. From the geometry of our sample, we estimate $E_\mathrm{c}\approx70~$mK, substantially lower than the charging energy ($\sim300~$mK) reported in the first observation of this effect. This is mainly due to the fact that in our experiment, the metallic island is extremely close ($15~$nm) to the graphite back gate, largely increasing its capacitance with respect to the experiment of ref.~\cite{Sivre2018}. Using this value, we compare the theoretical predictions to our experiment; unsurprisingly, the small value of the charging energy is such that the effect of heat Coulomb blockade is very weak. This is illustrated in Fig.~\ref{figsm_heatCBnu3-CD2}, showing the $\nu=3$ data for cooldown 2 along with its heat transport fit with $N=3$ fixed which yields $T_0=12.5~$mK, and the heat Coulomb blockade model with $T_0=15~$mK. The latter model reproduces slightly better our data; however, the difference between the two models remains at the limit of our experimental accuracy. Given this small difference, we neglect the effect of heat Coulomb blockade in our analysis.

\begin{figure}[ht]
\centering
\includegraphics[width=0.41\textwidth]{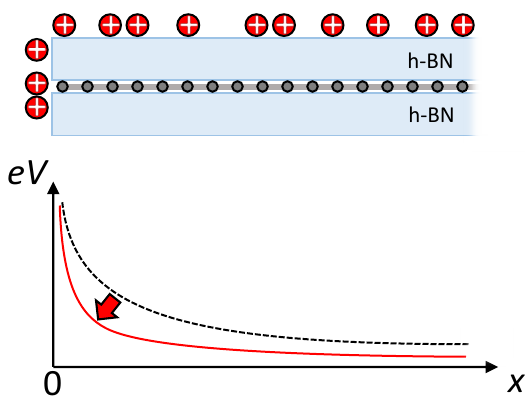}
\caption{\label{figsm_electrostatics} Top: side-view schematics (not to scale) of the h-BN/graphene/h-BN stack, with positively charged impurities (red balls) randomly scattered at its surface and on its edges in cooldown 2. Bottom: sketch of the electrostatic potential profile along the edge in absence of gate voltage. The black dashed line symbolizes the configuration of cooldown 1, and the red line the one of cooldown 2.}
\end{figure}

\subsection{Bulk heat transport and edge electrostatics}

We argue here that the observed decrease in the thermal conductance at $\nu=8/3$ in the second cooldown cannot stem from bulk contribution, despite the non-zero value of $\sigmaxx$. Indeed, electronic conduction in the bulk remains much smaller than on the edge, as shown by the fact that the reflected and transmitted transconductances are equal and reach their expected values (note also that the minimum of $\sigmaxx$ at $\nu=8/3$ corresponds to less than $1~\%$ of the electronic current carried through the bulk). The associated electronic bulk heat conduction should thus be negligible. Chargeless heat transport through the bulk cannot account either for the large change in $N$ observed in the second cooldown: the most drastic case where the upstream neutral edge mode is completely reflected across the bulk yields at best $N=3$, far beyond our uncertainty. Besides, the geometry of our sample, where the distance between the two edges is twice larger than the distance between the floating island and the nearest cold electrodes, is such that parasitic bulk heat conduction (be it charged or neutral) would effectively increase $N$ rather than diminish it. 

Fig.~\ref{figsm_electrostatics} shows a schematic representation of the charged impurities randomly adsorbed on the sample's surface during thermal cycling between cooldowns 1 and 2. As seen from the conductance data shown in main text Fig. 2, these impurities lead to an average increase of the intrinsic electron doping, such that they should be essentially positively charged. The positive charges adsorbed near the sides of the sample lower the electrostatic potential close to the edge, making the confinement potential sharper (red line in bottom graph of Fig.~\ref{figsm_electrostatics}). In heterostructures such as ours where the graphene is separated from the graphite gate by the encapsulating h-BN, several works \cite{Ribeiro-Palau2019,Flor2022} suggest that the typical length scale for the confinement potential is given by the thickness of the h-BN layer, that is 10-50 nm. Calculating the precise confinement potential as well as the lateral position of the edge channels in a graphene heterostructure has been the subject of very recent studies; in particular, a recent work \cite{Flor2022} shows that adding a local gate near the edge of a graphene/hBN heterostructures decreases the distance between edge channels by more than a factor 2. More precisely, they show that adding a local gate at 30 nm above a graphene flake that itself sits 30 nm above a global gate causes the distance between the nu=1 and nu=2 edge channels to decrease from about 70 nm to less than 30 nm. Even though this work does not exactly correspond to the geometry discussed in our work, there are enough similarities between the two (in particular the hBN thickness) to provide a reasonable estimate of the edge confinement.

\section{Cooldown 3}

\begin{figure}[ht]
\centering
\includegraphics[width=0.71\textwidth]{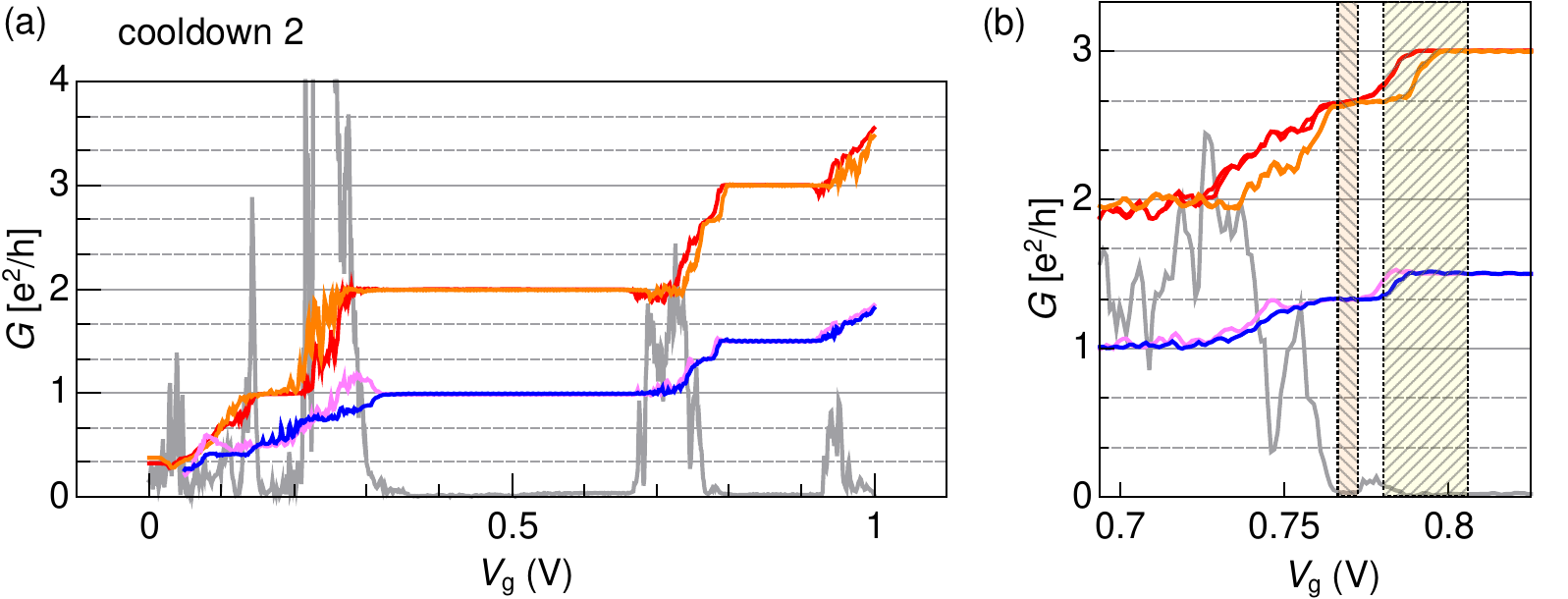}
\caption{\label{figsup_GvsVgcd3} \textbf{a,} Conductances versus gate voltage $\Vg$, measured at $B=7~$T and $T=12~$mK for cooldown 3. Red (resp. orange): 2-point conductance $G_{2\mathrm{pt}}$ of the transmitted (resp. reflected) side. Grey: longitudinal-like conductivity $\sigmaxx$. Lavender: reflected transconductance $G_\mathrm{R}$. Blue: transmitted transconductance $G_\mathrm{T}$. 
\textbf{b:} Zoom on the $\nu=2\rightarrow 3$ transition. The position of the $\nu=8/3$ plateau is indicated by the yellow shaded region for cooldown 1, and the orange shaded region for cooldown 2.}
\end{figure}

To investigate the reproducibility of our results, in particular the evolution of the conductance features, we have performed a third cooldown of the same sample after having kept it for several months in a dry atmosphere. We give here a brief summary of the results of this cooldown. The conductance versus $\Vg$ data obtained at $B=7~$T, and shown in Fig.~\ref{figsup_GvsVgcd3}, displays generally similar features as in cooldown 2. In particular, the position of the QH plateaus is the same, with an average gate voltage shift of about $-26~$mV with respect to cooldown 1 (as a reminder, this shif was equal to $-23~$mV for cooldown 2. The size of the $\nu=8/3$ plateau is also comparable to that of cooldown 2, although suprisingly slightly larger.

\begin{figure}[ht]
\centering
\includegraphics[width=0.3\textwidth]{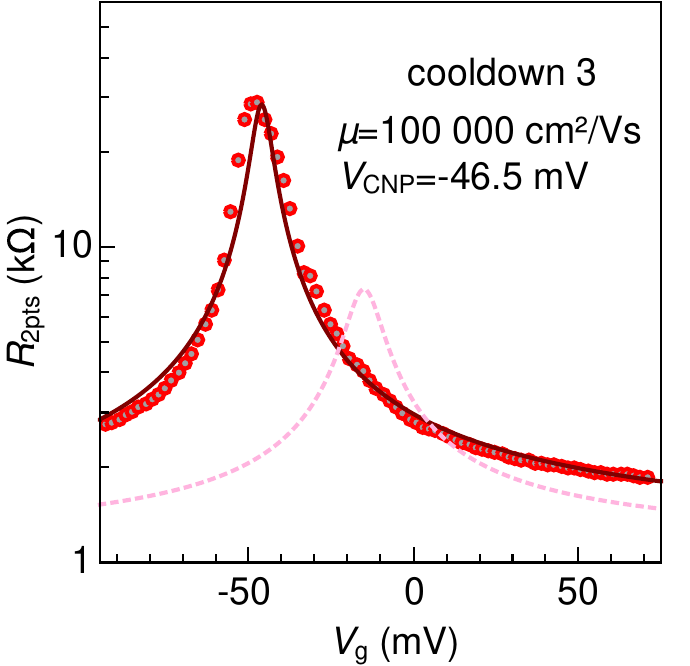}
\caption{\label{figsup_CNPcd3} Measurement of the 2-point resistance on the T side of the sample as a function of $\Vg$, at $B=0~$T and $T=12~$mK. Circles: data, dark red line: fit. The pink dashed line corresponds to the fit of the data at cooldown 1 shown in Fig.~\ref{figsup-CNPcd1}.}
\end{figure}

The resistance versus gate voltage data at zero magnetic field is shown in Fig.~\ref{figsup_CNPcd3}. It clearly displays a shift of the CNP towards negative $\Vg$. A fit of the data using the same equation as in Fig.~\ref{figsup-CNPcd1} yields a mobility of about $100000~$cm$^2/$Vs, indeed significantly smaller than at cooldown 1. The $\Vg$ position of the CNP is $-46.5~$mV, corresponding to a gate voltage shift of $-31~$mV that is comparable to the $-26~$mV $\Vg$ shift of the QH plateaus determined above.

\begin{figure}[ht]
\centering
\includegraphics[width=0.61\textwidth]{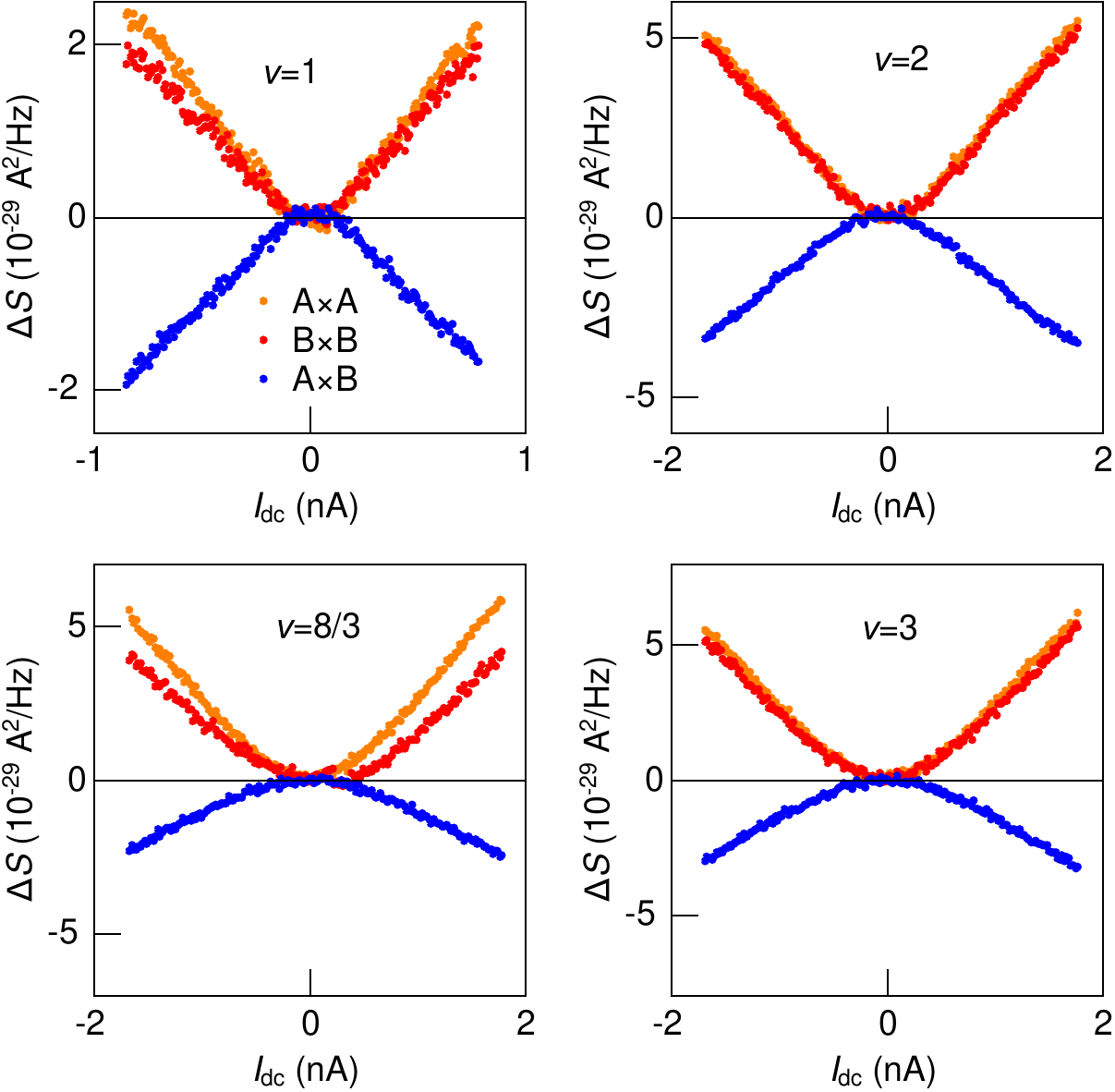}
\caption{\label{figsup_SvsIcd3} Auto- (orange and red symbols) and crosscorrelation (blue symbols) as a function of $\Idc$, for cooldown 3 at filling factors $\nu=1$ (top left), $2$ (top right), $8/3$ (bottom left), and $3$ (bottom right).}
\end{figure}

\begin{figure}[ht]
\centering
\includegraphics[width=0.45\textwidth]{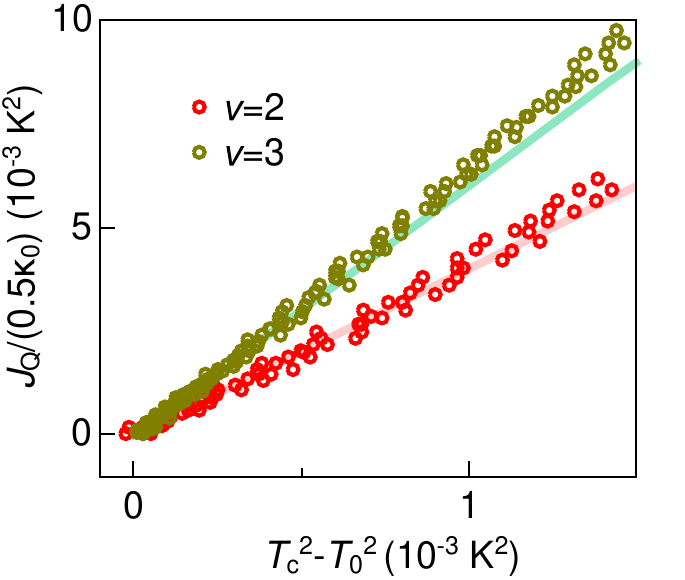}
\caption{\label{figsup_JqvsT2cd3} Heat flow $\Jq$ in units of $\kappa_0/2=\pi^2 k_{\mathrm{B}}^2/6h$ versus $\Tc^2-T_0^2$, for cooldown 3. Symbols: experimental data: red $\circ$~: $\nu=2$ and dark yellow $\circ$~: $\nu=3$). Lines: theoretical predictions for the quantized heat flow carried by $2N$ ballistic channels, with $N=2$ (pink) and $N=3$ (light green).}
\end{figure}

We have performed heat transport measurement with the procedure described in the main text and in the sections above. Fig.~\ref{figsup_SvsIcd3} shows measurements of the auto and cross-correlations as a function of $\Idc$ for cooldown 3, at various filling factor. While both autocorrelations are equal for $\nu=2$ and $3$, we found them to be significantly different for $\nu=1$ and $8/3$, indicating the presence of spurious noise contributions that cannot be easily subracted (see above). We relate this to an imperfect quantization of the reflected and transmitted transconductances at $\nu=1$ (see Fig.~\ref{figsup_GvsVgcd3}a), and to the fact that the $\nu=8/3$ plateaus for both sides are not identical in $\Vg$. Analyzing the $\nu=2$ and $3$ data in terms of heat flow yields the correct quantization shown in Fig.~\ref{figsup_JqvsT2cd3}, with a base electron temperature $T_0\approx27~$mK at $\nu=2$ and $T_0\approx24~$mK at $\nu=3$.

%bad noise at nu=1 and nu=8/3, OK at 2 and 3, with base temp 27/24 mK
